\documentclass[]{iopart} %iopart for JSTAT
\usepackage{hyperref}
\usepackage{bm}
\usepackage{graphicx}
\usepackage{color}
\makeatletter
\let\O\@undefined
\def\O{\mathrm{O}}

\newcommand{\nc}{\newcommand}

\eqnobysec %to obtain per section equation numbering with iopart
\nc{\grad}{\bigtriangledown}

\nc{\CA}{{\cal A}} \nc{\CB}{{\cal B}} \nc{\CC}{{\cal C}}
\nc{\CD}{{\cal D}} \nc{\CE}{{\cal E}} \nc{\CF}{{\cal F}}
\nc{\CG}{{\cal G}} \nc{\CH}{{\cal H}} \nc{\CI}{{\cal I}}
\nc{\CJ}{{\cal J}} \nc{\CK}{{\cal K}} \nc{\CL}{{\cal L}}
\nc{\CM}{{\cal M}} \nc{\CN}{{\cal N}} \nc{\CO}{{\cal O}}
\nc{\CP}{{\cal P}} \nc{\CQ}{{\cal Q}} \nc{\CR}{{\cal R}}
\nc{\CS}{{\cal S}} \nc{\CT}{{\cal T}} \nc{\CU}{{\cal U}}
\nc{\CV}{{\cal V}} \nc{\CW}{{\cal W}} \nc{\CX}{{\cal X}}
\nc{\CY}{{\cal Y}} \nc{\CZ}{{\cal Z}}

\nc{\bA}{\mathbb{A}} \nc{\bB}{\mathbb{B}} \nc{\bC}{\mathbb{C}}
\nc{\bD}{\mathbb{D}} \nc{\bE}{\mathbb{E}} \nc{\bF}{\mathbb{F}}
\nc{\bG}{\mathbb{G}} \nc{\bH}{\mathbb{H}} \nc{\bI}{\mathbb{I}}
\nc{\bJ}{\mathbb{J}} \nc{\bK}{\mathbb{K}} \nc{\bL}{\mathbb{L}}
\nc{\bM}{\mathbb{M}} \nc{\bN}{\mathbb{N}} \nc{\bO}{\mathbb{O}}
\nc{\bP}{\mathbb{P}} \nc{\bQ}{\mathbb{Q}} \nc{\bR}{\mathbb{R}}
\nc{\bS}{\mathbb{S}} \nc{\bT}{\mathbb{T}} \nc{\bU}{\mathbb{U}}
\nc{\bV}{\mathbb{V}} \nc{\bW}{\mathbb{W}} \nc{\bX}{\mathbb{X}}
\nc{\bZ}{\mathbb{Z}}

\newcommand{\rk}{\sqrt{\kappa}}
\newcommand{\rbar}{\bar{r}}

\newcommand{\vx}{{\bm x}}
\newcommand{\vy}{{\bm y}}

\newcommand{\comments}[1]{}

\newcommand{\vr}{{\bm r}}
\newcommand{\vk}{{\bm k}}
\newcommand{\vl}{{\bm l}}
\newcommand{\vq}{{\bm q}}
\newcommand{\vp}{{\bm p}}
\newcommand{\vP}{{\bm P}}
\newcommand{\vpb}{{\bar{\bm p}}}

\nc{\measure}{\frac{1}{\beta}\sum_{\omega_n} \int \frac{d^2 \vq}{(2\pi)^2}}
\newcommand{\meas}[2]{\frac{1}{\beta}\sum_{#1_n} \int \frac{d^2 {\bf #2}}{(2\pi)^2}}

\begin{document}

\title{Scrambling in the Quantum Lifshitz Model}
\author{Eugeniu Plamadeala and Eduardo Fradkin}
\address{Department of Physics and Institute for Condensed Matter Theory, University of Illinois at Urbana-Champaign, 1110 West Green Street, Urbana, Illinois 61801-3080, USA}
\date{\today}
\begin{abstract}
We study signatures of chaos in the quantum Lifshitz model through out-of-time ordered correlators (OTOC) of current operators. This model is a free scalar field theory with dynamical critical exponent $z=2$. It describes the quantum phase transition in  2D systems, such as quantum dimer models,  between a phase with an uniform ground state to another one with a spontaneously translation invariance. At the lowest temperatures the chaotic dynamics are dominated by a marginally irrelevant operator which induces a temperature dependent stiffness term. The numerical computations of OTOC exhibit a non-zero Lyapunov exponent (LE) in a wide range of temperatures and interaction strengths. The LE (in units of temperature) is a weakly temperature-dependent function; it vanishes at weak interaction and saturates for strong interaction. The Butterfly velocity increases monotonically with interaction strength in the studied region while remaining smaller than the interaction-induced velocity/stiffness.
\end{abstract}
\maketitle

\section{Introduction}
Scrambling is the delocalization of initially local quantum information over the entire system. It implies that information about the initial state cannot be deduced by any local measurements of the final state. This notion is closely related the notion of quantum chaos in many-body systems\cite{Hosur2016}. 
The so-called out-of-time-order correlators were invented to capture the failure of semiclassical methods in a study of superconductivity.\cite{Larkin1969} 
More recently it has been realized that they are a way to quantify chaos and information scrambling in black holes
\cite{KitaevPrize,ShenkerStanford2014,Maldacena2016}. 
Specifically, for generic local Hermitean operators $W$  and $V$, the signature of chaos is a regime of exponential growth of the OTOC $C(t) \sim e^{\lambda t}$ after local relaxation has occured but before the scrambling time when $C(t) \sim 1$. The  exponent $\lambda$ is called a ``Lyapunov exponent'' by analogy with the 
Lyapunov exponents of classical dynamical systems:\cite{Arnold-1978,Moser-1973} in both classical and quantum cases the system looses memory 
of its initial state exponentially fast. It is generally conjectured that systems with this behavior are ``chaotic''. 
In the classical setting this means that the system over very long times eventually explores all its available phase state with probability 1. 
Although the precise definition of chaos is lacking (or rather there is an abundance of them\cite{Hosur2016}), in the quantum setting of a chaotic system it is believed that, at times long compared with the 
scrambling time, the system becomes described by a density matrix and in some sense has thermalized.\cite{Sekino-2008} 
The out-of-time ordered correlator $C(t)$ (OTOC) of two hermitian operators $V$ and $W$ is
%. is hypothesized to equal the Lyapunov exponent characterizing chaos of the classical limit of the theory (although there is evidence to the contrary\cite{Galitski2017}).
%{\bf TODO: most solid evidence that non-zero chaos exponent implies some kind of "quantum chaos"?}
\begin{eqnarray}
\label{eqn:general-otoc}
	C(t) &= \tr \lbrace \rho  [W(t), V]^\dagger [W(t),V] \rbrace
\end{eqnarray}
Numerous works have recently studied quantum chaos both at strong\cite{SY1993,KitaevPrize} and weak
 coupling,\cite{Stanford2016,ChowdhurySwingle2017} in gapped phases, in rational CFTs,\cite{RobertsCFT2015, Gu2016} 
 and in strongly disordered phases.\cite{MBL1,MBL2,ChowdhurySwingleDisordered} 
 
In this paper we will compute the chaos (Lyapunov)  exponents 
 and the Butterfly  velocity for a system close to  a quantum critical point. In general, this is a challenging problem since most quantum 
 critical points are associated to a non-trivial UV fixed point of a quantum field theory. However in 2+1 dimensions there is a quantum critical theory, 
 the quantum Lifshitz model \cite{ardonne-2004} (qLM), which is a free (compactified) scalar field theory with dynamical exponent $z=2$. 
 Since this fixed point theory is a free field, it is not chaotic. In spite of its simplicity, the qLM is a non trivial theory with a rich spectrum of operators, 
 and has been studied quite extensively.\cite{ardonne-2004,Fradkin-2004,Stephan-2009} 
 One aspect of this theory that is useful in this context is that it has a marginally irrelevant operator\cite{Fradkin-2004,Ghaemi-2005,Hsu-2013} 
 which, if included in the Lagrangian, spoils the integrability of the fixed point theory. 
 This is thus a controlled setting to examine the behavior of the OTOCs of this theory within a perturbative approach.

The   imaginary-time action of the qLM in 2+1-dimensions is:
\begin{eqnarray}
	S_0 &= \frac{1}{2}  \int d\tau d^2x \left ( \left (\partial_\tau \phi \right )^2  + \kappa \left ( \grad^2 \phi \right )^2 \right ) 
\end{eqnarray}

There are several variations of this model: the compactified and non-compactified real scalar, as well as the complex scalar.
The leading irrelevant operator at all these critical points is the same - $\left ( \grad \phi \right )^4$, and for this reason we believe the physics 
concerning chaos will be {\it qualitatively} the same. For simplicity, here we choose to study the case of the non-compactified real scalar theory.

Since this QCP is described by a quadratic action, it is integrable and cannot thermalize. 
Lack of integrability is not sufficient to guarantee thermalization.\cite{Gogolin2011}
The $\left ( \grad \phi \right )^4$ interaction is meant to both break integrability\cite{Prosen1993} and lead to thermalization, 
although we have no proof it does this. It is marginally irrelevant at the QCP and, barring $\cos(n\phi)$ terms only allowed 
(but less relevant at $\kappa \ll 1$) in the compactified theory, it is the the leading perturbation to the scaling form of the action. 
As with other irrelevant operators, its main effect  is to shift the location of the QCP by inducing a  renormalized stiffness, 
the coefficient of the relevant operator $(\grad \phi)^2$, whose renormalized value is tuned to zero at $T=0$ (but {\it not} at $T>0$) as the 
definition of the QCP.  Thus, our approach consists of   deforming $S_0$ by the following:
\begin{eqnarray}
	S' &=  \int d\tau d^2x  \left [ r_{qc} \left ( \grad \phi \right )^2 + \frac{u}{4} \left (\grad \phi \right )^4 \right ]
\end{eqnarray}
where the stiffness $r_{qc} = r_{qc}(u)$ is chosen such that we hit the QCP as $T \rightarrow 0$ keeping the 
bare value of $u$ fixed. The field $\phi$ is non-compact.

In order to perform a controlled calculation we choose to consider an $N$-flavor version of this model:

\begin{eqnarray}
\fl	\CL_{QLM,N} = \sum_a \frac{1}{2}\left [ (\partial_\tau \phi_a)^2 + r_{qc} (\grad \phi_a)^2 + \kappa (\grad^2 \phi_a)^2 \right ] 
	+ \frac{u}{4N}\sum_{a,b}(\grad \phi_a)^2 (\grad \phi_b)^2 
\end{eqnarray}
Furthermore it is helpful to rewrite the $(\grad \phi)^4$ term by introducing an auxiliary field $\sigma$ through a Hubbard-Stratonovich transformation:
\begin{equation}
	Z_E = \int \CD \phi e^{-S_{QLM,N}[\phi]}
		= \int \CD \phi \CD\sigma e^{-S_{N}[\phi,\sigma]}
\end{equation}
In this form the (Euclidean) Lagrangian is
\begin{eqnarray}
\fl \CL_N[\phi,\sigma] = \sum_a \frac{1}{2}\left[ (\partial_\tau \phi_a)^2 + r_{qc} (\grad \phi_a)^2 + \kappa (\grad^2 \phi_a)^2 \right] 
- \frac{\sigma^2}{4u} + \frac{\sigma}{2\sqrt{N}} \sum_a \left ( \grad \phi_a \right )^2
\end{eqnarray}

As discussed in Appendix~\ref{app:saddle-point-solution}, at $N=\infty$ we can solve the theory most naturally by introducting a new coupling:
%$\frac{\sigma}{\sqrt{N}} + r_{qc} = r_{eff}$ where 
\begin{equation}
r_{\rm eff}=\frac{\langle \sigma \rangle}{\sqrt{N}}+r_{qc} \quad \textrm{where} 	\quad r_{\rm eff} = \frac{uT}{4\pi \kappa} \frac{\ln\left ( \frac{\rk T}{r_{\rm eff}}\right )}{1+\frac{u}{16\pi\kappa^{3/2}}\left [ \ln\left (\frac{4 \kappa \Lambda^2}{r_{\rm eff}}\right )\right ]}
\end{equation}
We note that the expectation value of the auxiliary field $\langle \sigma \rangle$ induces a stiffness (``velocity'') for the $\phi$ field (instead of a ``thermal mass" as in  $\phi^4$ theory). We now discuss solutions of the equation above. \newline
At fixed interaction strength $u$ and sufficiently low temperatures the logarithm dominates the denominator and we obtain (see discussion around Eqn\ref{app:reff-equation})
\begin{eqnarray}
\label{eqn:reff-solution}
	r_{\rm eff} & \approx 4 T \rk \frac{\ln\ln \left (\frac{\Lambda^2}{T \rk}\right )}{\ln \left (\frac{\Lambda^2}{T\rk}\right )}
\end{eqnarray}
which is consistent with our assumptions at low $T$. The marginality of the $(\grad \phi)^4$ term has induced log corrections that violate $z=2$ scaling. 
Note that $r_{\rm eff}$ appears to be independent of the interaction strength $u$ at low $T$. This is only true when the approximation we made in the denominator holds: $\frac{u}{16\pi\kappa^{3/2}}\left [ \ln\left (\frac{4 \kappa \Lambda^2}{r_{\rm eff}}\right )\right ] \gg 1$. The assumption we began with is clearly valid. We will call this the {\it low temperature regime}.\newline
On the other hand at higher temperatures or small $u$ the opposite is true
\begin{eqnarray}
	\frac{u}{16\pi\kappa^{3/2}}\left [ \ln\left (\frac{4 \kappa \Lambda^2}{r_{\rm eff}}\right )\right ] \ll 1
\end{eqnarray}
and using similar methods we obtain a different saddle-point equation
\begin{eqnarray}
\fl	r_{\rm eff} = \frac{uT}{4\pi \kappa} {\ln\left ( \frac{\rk T}{r_{\rm eff}}\right )} \quad \textrm{with solution} \quad 	r_{\rm eff} = \sqrt{\kappa} T \left (\frac{u}{4\pi \kappa^{3/2}}\right ) \ln \left (\frac{4\pi \kappa^{3/2}}{u} \right )
\end{eqnarray}
We call this the {\it weak interaction regime}. We will study both regimes simultaneously by definig $\alpha(u,T) = r_{\rm eff}(u,T)/T$. Since $\alpha$ has very weak temperature dependence in both regimes, we will often treat $\alpha$ as an indepedent parameter (in a wide but finite range of temperatures). \newline

%%%%%

The $1/N$ corrections are then captured by the following action (see \ref{app:1-n-corrections}) where the $\lambda$ field is the deviation from the saddle-point value of $\sigma$, i.e. $\lambda=\sigma-\langle \sigma \rangle_{N=\infty}$,
\begin{eqnarray}
\label{eqn:phi-lambda-N}
\fl	\CL_{\phi,\lambda,N} = \sum_a \frac{1}{2}\left[ (\partial_\tau \phi_a)^2 + r_{\rm eff} (\grad \phi_a)^2 + \kappa (\grad^2 \phi_a)^2 \right] - 
	\frac{\lambda^2}{4u} + \frac{\lambda}{2\sqrt{N}} \sum_a \left( \grad \phi_a \right)^2
\end{eqnarray}
At leading order in $N$ the $\phi_a$ fields decouple from $\lambda$ and each other, and there is no chaos. 
Therefore, the chaos (Lyapunov) exponent is at most of order $1/N$ at large $N$ and we must study $1/N$ corrections using the theory 
of Eq.\ref{eqn:phi-lambda-N}.

The building blocks of our perturbative expansions are the imaginary-time $\phi_a$ propagator
\begin{eqnarray}
	\CG(\tau,\vx) \delta_{a,b} &= \langle T_\tau \phi_a(\tau,\vx) \phi_b(0,0) \rangle \\
	\CG(i\omega_n, \vq) &= \frac{1}{\omega_n^2 + \epsilon_\vq^2}, \quad \textrm{ where } \quad \epsilon_\vq^2 = r_{\rm eff}(u,T) \vq^2 + \kappa \vq^4
\end{eqnarray}

The retarded $\phi_a$ propagator $\CG_R(\omega,\vk)$ is obtained as usual by analytic continuation $i\omega_n \rightarrow \omega + i0$. The spectral function  is
\begin{eqnarray}
	A(\omega,\vk) &= -2 \Im \CG_R(\omega,\vk) =  \frac{\pi}{\epsilon_\vk} \left [\delta(\omega-\epsilon_\vk) - \delta(\omega+\epsilon_\vk) \right ]
\end{eqnarray}
We will also make use of the symmetrized Wightman propagator of the $\phi$ fields, defined as
\begin{eqnarray}
	\CG_W(t,\vx)\delta_{a,b} &= \tr \lbrace \rho^{1/} \phi_a(\vx,t) \rho^{1/2} \phi_b(0) \rbrace
\end{eqnarray}
By manipulations similar to those in Appendix C of Ref.[\cite{ChowdhurySwingle2017}] the Wightman propagator can be related to the spectral function through
\begin{eqnarray}
	\label{eq:phi-wightman}
	\CG_W(\omega,\vk) &= \frac{A(\omega,\vk)}{2\sinh \frac{\beta \omega}{2}}
\end{eqnarray}

We will shortly need the $\phi$ and $\lambda$ propagators dressed to leading order. We denote the bare $\lambda$ field propagator as $\CG^{0}_\lambda(\tau,\vx) = -2u$. It picks up a $O(N^0)$ correction through its self-energy $\Pi(i\omega_n,\vq)$ to give the dressed $\lambda$ propagator
\begin{eqnarray}
	\CG_\lambda(i\omega_n,\vq) = \frac{1}{\frac{1}{-2u} - \Pi(i\omega_n,\vq)}
\end{eqnarray} 
from which one can derive the other functions: 
\begin{eqnarray}
	\label{eq:lambda-retarded-propagator}
\fl	 \textrm{the retarded propagator} & \quad  \CG_{R,\lambda}(\omega,\vq) &= -\CG_\lambda(i\omega_n \rightarrow \omega + i0,\vq)\\
\fl	 \textrm{the spectral function} & \quad A_\lambda(\omega,\vq) &= -2 \Im \CG_{R,\lambda}(\omega,\vq) \\
\fl	 \textrm{the Wightman propagator} & \quad   \CG_{W,\lambda}(\omega,\vq) &= \frac{A_\lambda(\omega,\vq)}{2\sinh \frac{\beta \omega}{2}}
\end{eqnarray}

The Feynman rule for the $\phi-\lambda$ vertex follows from
\begin{eqnarray}
\fl	-\CL_E \propto \frac{1}{2\sqrt{N}} \int_{\omega_n,\omega_m} \int_{\vk,\vq} (-\vk \cdot \vq) \lambda(-i\omega_n-i\omega_m, -\vk-\vq) \phi_a(i\omega_n,\vk) \phi_a(i\omega_m,\vq)
\end{eqnarray}
Here and below we use the notation $\int_\vp = \int \frac{d^2 p}{(2\pi)^2}$ and $\int_{\omega_n} = T\sum_{\omega_n}$. The vertex then equals $\frac{-\vk_1 \cdot \vk_2}{\sqrt{N}}\delta(\sum_i \vk_i)$, where the dotted momenta belong to the $\phi$ fields.

\begin{figure}[hbt]
\begin{center}
	\includegraphics[width=0.5\textwidth]{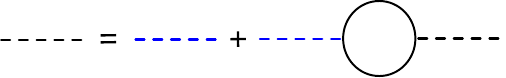}
	\caption{The self-energy correction of the $\lambda$ field at leading order in $1/N$.}
	\end{center}
	\label{fig:pi-one-loop}
\end{figure}
The one-loop self-energy $\Pi(i\omega_n, \vq)$ of the $\lambda$ field comes from the Dyson equation for the field $\lambda$, the Feynman diagram 
shown in Fig.\ref{fig:pi-one-loop} and equals
\begin{eqnarray}
	\Pi(i\omega_n,\vq) &= \frac{T}{2}\sum_{i\nu_n} \int_\vk^\Lambda \frac{\left ( \vk \cdot (\vk + \vq) \right )^2 }{\left( \nu_n+ \omega_n \right )^2 + \epsilon^2_{\vk+\vq} } \frac{1}{\nu_n^2 + \epsilon^2_\vk} 
\end{eqnarray}
where the factor of $1/2$ comes from combinatorics ($1/2$ from second order of exponential series, and two factors from vertex; on top there 
are two ways to contract the $\phi$'s, and two to contract the $\lambda$'s). In \ref{app:lambda-self-energy} we discuss the computation 
of this object, which must ultimately be done numerically.

%\begin{figure}[hbt]
%\begin{center}
%	\includegraphics[width=0.5\textwidth]{phi_selfenergy.pdf}
%	\caption{The leading contributions to the self-energy of the $\phi$ field. The second diagram induces only a real shift of the pole and plays no role in our calculation.}
%	\end{center}
%	\label{fig:sigma-one-loop}
%\end{figure}

Similarly, the one-loop self-energy of $\phi$ is given by the Feynman diagrams:
\begin{equation}
\label{eq:sigma-one-loop}
\begin{minipage}[b]{0.18\textwidth}
	$\Sigma(i\omega_n,\vq)=$ \\ \vspace{10pt}
\end{minipage}
\begin{minipage}[c]{0.5\textwidth}
\includegraphics[width=\textwidth]{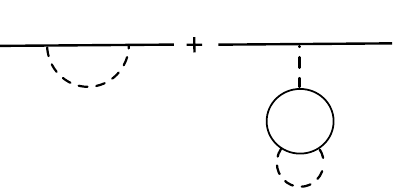}
\end{minipage}
\end{equation}
The second diagram in Eq.(\ref{eq:sigma-one-loop}) induces only a real shift of the pole and plays no role in our calculation. 
\begin{eqnarray}
\fl	\Sigma\left (i\omega_n,\vq \right ) &= \frac{T}{N}\sum_{ i \nu_n} \int_\vk^\Lambda \left ( \vq \cdot (\vq+\vk) \right )^2 
	\CG(i\omega_n + i\nu_n, \vq+\vk) \CG_\lambda(i\nu_n,\vk)
\end{eqnarray}
where $\CG_\lambda$ is already one-loop corrected. This object is computed in \ref{app:phi-scattering-rate}.
Note that the second diagram also contributes at order $1/N$ but as it induces only a real shift of the pole we drop it with the 
philosophy that the location of the pole is a physical quantity that we choose to keep fixed with a suitable counterterm.

\section{Perturbative expansion of OTOC}

\begin{figure}[hbt]
\begin{center}
	\includegraphics[width=0.2\textwidth]{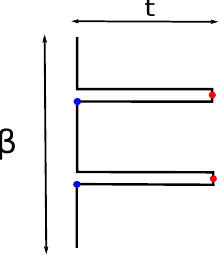}
	\caption{Complex time contour of the operators insertions of the regulated OTOC. The blue (red) dot corresponds to the $V(0)$ ($W(t)$) operator.}
	\end{center}
	\label{fig:time-contour}
\end{figure}

Instead of studying the OTOC of Eq.\ref{eqn:general-otoc} we will study a closely related object with the same growth properties. 
Following the approach employed for matrix $\phi^4$ theory in the large $N$ limit by Stanford\cite{Stanford2016}, we shift half the fields half-way along the thermal circle.\cite{ChowdhurySwingle2017} 
This removes spurious short-distance divergences of coincident operators, without modifying the exponential growth.
% {\bf TODO: convince myself of this}.
The resulting operators insertions in complex time lie on the contour shown in Fig.~\ref{fig:time-contour}.

Specifically we choose to study the following (regulated) OTOC
\begin{eqnarray}
\fl	C(t_1-t_2,\vx_1 - \vx_2) = \nonumber \\
\fl \qquad -\frac{1}{N^2}\sum_{a,b}{\rm Tr} \lbrace \sqrt{\rho} \left [ \grad \phi_a(\vx_1,t_1), \grad \phi_b(\vx_2,t_2) \right ] \sqrt{\rho} 
	\left[ \grad \phi_a(\vx_1,t_1), \grad \phi_b(\vx_2,t_2) \right] \rbrace
\end{eqnarray}
Note that gradients are dotted into each other within each commutator. The OTOC can be expanded in powers of the interaction vertex.

We derive the rules in \ref{app:oto-rules} (they are almost identical to those in Chowdhury and Swingle \cite{ChowdhurySwingle2017}), 
and simply present them here:
\begin{enumerate}
	\item Vertex insertions can occur on either time-fold. Each comes with a factor of $\frac{-i \vk_1 \cdot \vk_2}{2\sqrt{N}}\delta_{ab}$ 
	(coming from $\phi_a,\phi_b$). Contractions inside each time fold represent self-energy corrections and are counted separately. 
	\item Horizontal lines correspond to retarded propagators $i\CG_R, i\CG_{R,\lambda}$, vertical lines to Wightman propagators $G_W, G_{W,\lambda}$.
	\item Each line must be directed, left to right or top to down. This determines whether a momentum/frequency is incoming or outgoing. 
	The sum of incoming must equal sum of outgoing at each vertex.
\end{enumerate}

\begin{figure}[hbt]
\begin{center}
	\includegraphics[width=0.3\textwidth]{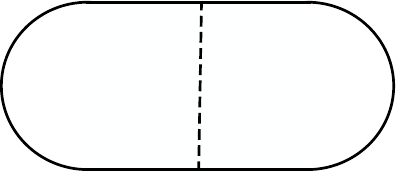}
	\caption{The simplest ladder diagram with rungs of type 1, $C_1(\nu,\vq=0)$ defined in Eq.\ref{eqn:rung1-expression}.}
	\end{center}
	\label{fig:type-1-example}
\end{figure}

There are two classes of diagrams then that are most important, both involve contractions between the different time folds.
The first class involve ladder rungs of $\lambda$ propagators (we will call them type 1). The simplest example of a type 1 diagram is shown in Fig.\ref{fig:type-1-example}.

Its value at $\vq=0$ momentum transfer is %(see Eq.\ref{eqn:rung1-expression} for full derivation):\\
%\textcolor{blue}{\bf EF: Where is this derived?}
\begin{eqnarray}
\label{eqn:rung1-expression}
\fl	C_1(\nu,\vq=0) 
	= \frac{1}{N^2}  \int_{\vp,\vp'} \left ( \vp \cdot \vp'\right )^4 \int_{\omega,\omega'} \CG_R(\vp,\omega) & \CG_R(-\vp,\nu-\omega) \CG^W_\lambda(\vp-\vp',\omega-\omega')  \nonumber \\
	& \times  \CG_R(\vp',\omega')\CG_R(-\vp',\nu-\omega')
\end{eqnarray}
For clarity and conciseness, in Eq.\ref{eqn:rung1-expression} we present the expression at $\vq=0$. The full expression for $C_1(\nu,\vq) $ is worked out in \ref{app:oto-rules}.

\begin{figure}[hbt]
\begin{center}
	\includegraphics[width=0.3\textwidth]{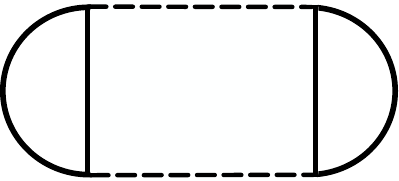}
	\caption{The simplest ladder diagram with rungs of type 2, $C_2(\nu,\vq=0)$ defined in Eq.\ref{eqn:rung2-expression}.}
	\end{center}
	\label{fig:type-2-example}
\end{figure}

An example of the second class of diagrams is shown in Fig.\ref{fig:type-2-example}. These involve ladder rungs of $\phi$ propagators (we will call them type 2).
The expression for  the simplest such diagram is
\begin{eqnarray}
	\label{eqn:rung2-expression}
\fl	C_2(\nu,\vq=0) = \frac{1}{N^2} \int_{\vp,\vp'} \left ( \vp \cdot \vp' \right )^2 \int_{\omega,\omega'}  \, 
	& \CG_R(\nu-\omega,-\vp) \CG_R(\omega,\vp)  \CG_{\rm eff}(\nu,\vq;\omega',\omega,\vp',\vp) \nonumber \\
	& \times \CG_R(\nu-\omega',-\vp') \CG_R(\omega',\vp') 
\end{eqnarray}
where $\CG_{\rm eff}$ above represents the loop of four propagators that form the inner rectangle in Fig~\ref{fig:type-2-example}. It is given by
\begin{eqnarray}
	\label{eqn:G-eff}
\fl	\CG_{\rm eff}(\nu,\vq=0;\omega',\omega,\vp',\vp) = \int_{\omega'',\vp''} & \left ( \vp \cdot (\vp''-\vp) \right )^2 
		 \left (  \vp' \cdot (\vp'-\vp'') \right )^2   \nonumber \\ 
		 	& \times \CG_W(\omega''-\omega,\vp''-\vp) \CG_W(\omega'-\omega'',\vp'-\vp'') \nonumber  \\
	& \times \CG_{R,\lambda}(\nu-\omega'',-\vp'') \CG_{R,\lambda}(\omega'',\vp'') 
\end{eqnarray}
The functions $\CG_W$ and $\CG_{R,\lambda}$ are defined in Eqns~\ref{eq:phi-wightman} and \ref{eq:lambda-retarded-propagator}. In the sequel, we will typically suppress the $\nu$ and $\vq$ dependence of $\CG_{\rm eff}$ and consider it implicit.

%TODO: argue away crossed diagrams

At late times both $C_1$ and $C_2$ contribute at order $t/N^2$. A general diagram with $n_1$ rungs of type 1 and $n_2$ rungs of type 2 is of order $1/N^{n_1+n_2+1}$, so we expect their sum to be of order $1/N$. We ignored crossed diagrams because they are parametrically smaller than the ladder diagrams (see \ref{app:crossed-diagrams}).

\section{Computation of the OTOC and the Bethe-Salpeter Equation}

Before we compute the OTOC, $\CC(t)=\int d^2x \; \CC(t,\vx)$,
we will briefly describe the structure of the calculation. \newline
Instead of aiming directly for $\CC(t,\vx)$ we will obtain $\CC(\nu,\vq)$, which is its Laplace transform in frequency and the Fourier transform in momentum. The $\vq=0$ value is sufficient to compute the Lyapunov exponent $\lambda_L$, while the $\vq \neq 0$ value is necessary to obtain the Butterfly velocity (and a semi-independent verification of $\lambda_L$). \newline

\begin{figure}[hbt]
\begin{center}
	\includegraphics[width=0.4\linewidth]{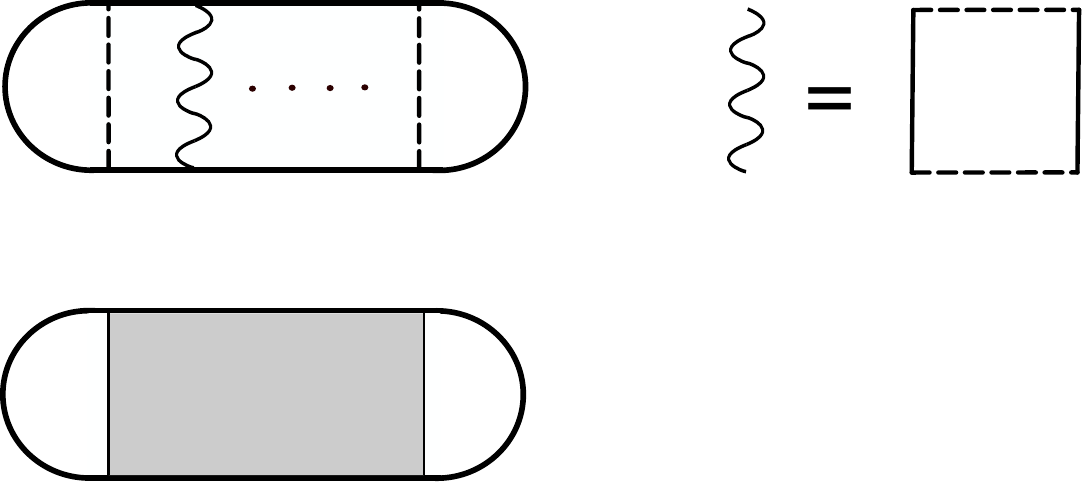}
	\caption{Top left is a general uncrossed diagram. The bottom represents the complete sum $\CC(\nu)$. }
	\end{center}
	\label{fig:general-ladder-diagram}
\end{figure}

The function $\CC(\nu,\vq)$ has a diagrammatic expansion in powers of $1/N$. The first two terms are $C_1(\nu,\vq)$ and $C_2(\nu,\vq)$ previously defined in Eqns~\ref{eqn:rung1-expression} and \ref{eqn:rung2-expression} corresponding to diagrams with rungs of type 1 and 2 respectively. We will set up and solve the Bethe-Salpeter equation for $\CC(\nu,\vq)$ in terms of an auxiliary function $g(\nu,\vq; \omega,\vp)$ defined by
\begin{eqnarray}
	\CC(\nu,\vq) &= \frac{1}{N} \int\frac{d\omega}{2\pi}\int_\vp \vp^2 g(\nu,\vq; \omega,\vp),
\end{eqnarray}
The Bethe-Salpeter equation will be finally recast to an integral equation, which we will need to solve numerically. To do that we will discretize and turn it into a matrix eigenvalue problem. Finally the eigenvalue with the largest positive real part will dominate the exponential growth of the OTOC in time.
\newline
We study the $\vq=0$ case first. We subsequently suppress the $\vq$ index (when zero) and write $\CC(\nu)$ in place of $\CC(\nu,\vq)$.
To compute the OTOC $\CC(\nu)$ we now proceed with an evaluation of the sum of ladder diagrams of Fig.\ref{fig:general-ladder-diagram}.

%\begin{wrapfigure}{r}{0.5\textwidth}%[hbt]
%	\includegraphics[width=0.7\linewidth]{bethe_salpeter.pdf}
%	\caption{A diagrammatic definition of the Bethe-Salpeter equation for $\CC(t)$.}
%	\label{fig:bethe-salpeter}
%\end{wrapfigure}

\begin{figure}[hbt]
\begin{center}
	\includegraphics[width=0.7\linewidth]{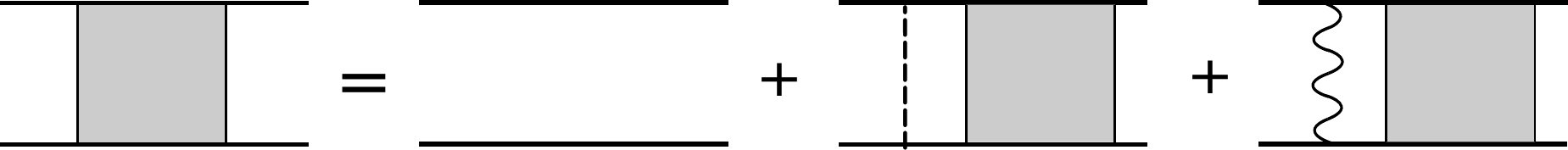}
	\caption{A diagrammatic definition of the Bethe-Salpeter equation for $\CC(t)$.}
	\end{center}
	\label{fig:bethe-salpeter}
\end{figure}

To this end we set up a Bethe-Salpeter equation, which diagrammatically is represented in figure Fig.\ref{fig:bethe-salpeter}.
We write down the value of the first few diagrams with rungs of type 1:
\begin{eqnarray}
\fl	\CC(\nu) &= \frac{1}{N}\int_{\omega,\vp} \vp^4 \CG_R(\nu-\omega,-\vp)\CG_R(\omega,\vp) \\
\fl	&+  \frac{1}{N^2}\int_{\omega,\omega',\vp,\vp'} \vp^2 \CG_R(\nu-\omega,-\vp)\CG_R(\omega,\vp) \left ( \vp \cdot \vp' \right )^2 
	\CG_{W,\lambda}(\omega'-\omega,\vp'-\vp) \nonumber \\
\fl	& \qquad  \qquad \times \CG_R(\nu-\omega',-\vp')\CG_R(\omega',\vp') \vp'^2 \\
\fl	&+\frac{1}{N^3} \int \vp^2 \left ( \vp \cdot \vp' \right )^2 \left ( \vp' \cdot \vp'' \right )^2 (\vp'')^2   G^2_\vp  \CG_{W,\lambda} G^2_{\vp'} 
	\CG_{W,\lambda} G^2_{\vp''} + \cdots
\end{eqnarray}
where the last line is written schematically. This allows us to infer the Bethe-Salpeter equation for $\CC(\nu)$, written in terms of the auxiliary function $g$:
\begin{eqnarray}
	\label{eq:bethe-salpeter-zeroq}
\fl	g(\nu; \omega,\vp) = &\CG_R(\nu-\omega,-\vp)\CG_R(\omega,\vp) \nonumber \\ 
	&\times \left [ \vp^2  + \frac{1}{N}\int_{\omega',\vp'} 
	\left ( \vp \cdot \vp' \right )^2 \CG_{W,\lambda}(\omega'-\omega,\vp'-\vp)  g(\nu;\omega',\vp') \right ]
\end{eqnarray}
We can also include the type 2 rungs  by making the following replacement in the expression given above
\begin{eqnarray*}
\fl	\left ( \vp \cdot \vp' \right )^2 \CG_{W,\lambda}(\omega'-\omega,\vp'-\vp)  \rightarrow \left ( \vp \cdot \vp' \right )^2 
	\CG_{W,\lambda}(\omega'-\omega,\vp'-\vp) + \CG_{\rm eff}(\nu,\vq=0;\omega',\omega; \vp',\vp)
\end{eqnarray*}
where $\CG_{\rm eff}$ is given by Eq.\ref{eqn:G-eff}.

To make progress and extract the leading time-dependence we must simplify the object $\CG_R(\nu-\omega,-\vp)\CG_R(\omega,\vp)$. We do this by the following replacement:
\begin{eqnarray}
		\CG_R(\nu-\omega,-\vp)\CG_R(\omega,\vp) & \rightarrow \frac{\pi i }{\epsilon_\vp} \frac{\delta(\omega^2 - \epsilon_\vp^2)}{\nu + i 2 \Gamma_\vp} 
\end{eqnarray}
As explained in \ref{app:product-retarded-simplification} this replacement achieves three things: it simplifies the expression, it captures the pole structure that leads to leading late-time dependence in $\CC(t)$, and finally it also captures the scattering rate, $\Gamma_\vp$, of the $\phi$ fields (see discussion around Eqn.\ref{eqn:supplement-simplifying-GR}). 

Since the product of the two Green functions $\CG_R$ has an on-shell delta function condition, we choose the following ansatz
\begin{eqnarray}
	g(\nu;\omega,\vp) &= g(\nu;\vp) \delta( \omega^2 - \epsilon_\vp^2 ) 
\end{eqnarray}
This simplifies the Bethe-Salpeter equation~\ref{eq:bethe-salpeter-zeroq} to the following 
\begin{eqnarray}
		\label{eqn:g-integral-eqn}
\fl		-i\nu \, g(\nu;\vp) &= \frac{\pi \vp^2}{ \epsilon_\vp} + \frac{1}{N}\int_{\vl} \left ( \CR_1(\vl,\vp) + \CR_2(\vl,\vp) - 2 N \Gamma_\vp \, \delta^{(2)}(\vl-\vp) \right ) g(\nu;\vl)
\end{eqnarray}
where we introduce auxiliary functions $\CR_1, \CR_2$ to compactify the expression. They are given by
\begin{eqnarray}	
	\CR_1(\vl,\vp) &= \CR_{1,+}(\vl,\vp) + \CR_{1,-}(\vl,\vp) \\
	\CR_{1,\pm}(\vl,\vp) &= \frac{\left (\vl \cdot \vp \right )^2}{4 \epsilon_\vl \epsilon_\vp} {\CG_{W,\lambda}(\pm \epsilon_\vl - \epsilon_\vp, \vl-\vp)} \\
	\CR_2(\vl,\vp) &= \CR_{2,+}(\vl,\vp) + \CR_{2,-}(\vl,\vp) \\
	\CR_{2,\pm}(\vl,\vp) &= \frac{1}{4 \epsilon_\vl \epsilon_\vp} \CG_{\rm eff} \left( \pm \epsilon_\vl, \epsilon_\vp; \vl-\vp, \frac{\vl+\vp}{2} \right)
\end{eqnarray}

Note that $\CR_{2,\pm}(\vl,\vp)$ actually depends on $\nu, \kappa, u, T$ in addition to $\vl,\vp$. Similarly $\CR_1$ depends on $\kappa, u, T$. We convinced ourselves numerically that indeed $R_1(\vl,\vl)$ vanishes. Both $R_1(\vl,\vp)$ and $R_2(\vl,\vp)$ also vanish when either $\vl,\vp \rightarrow 0$ due to the factors of momentum in the numerator.

\section{Lyapunov Exponent and Butterfly velocity}

We note that \ref{eqn:g-integral-eqn} can be made dimensionless by the replacement $\nu \rightarrow \nu T, \vp \rightarrow \vp \sqrt{T}, r_{\rm eff}(u,T) = \alpha(u,T) T$, with $[g] = [1/T] = 2$. Then, by dimensional considerations and inspection of $r_{\rm eff}$, the Lyapunov exponent $\lambda_L$ will take the form
\begin{equation}
\lambda_L = n(u,\alpha(u,T)) \frac{T}{N}
\label{eq:lambdaL}
\end{equation} 
where $n(u,\alpha(u,T))$ is a dimensionless function with the following behaviors
\begin{itemize}
	\item	In the {\it weak interaction regime} $n(u,T) = n(u)$ is temperature independent with $n(u=0) = 0$ (since at $u=0$ the model becomes integrable).
	\item	In the {\it low temperature regime}, $n(u,T)$ inherits a weak temperature dependence from $\alpha(T)$. 
\end{itemize}

We solve the integral equation of Eq.\ref{eqn:g-integral-eqn} numerically. Given the eigenfunction $g_\eta(\nu;\vp)$ of the right-hand-side integral 
with eigenvalue $\eta$ we obtain the following final form for \ref{eqn:g-integral-eqn}:
\begin{eqnarray}
	\label{eqn:g-eigenfunction}
	-i\nu \, g_\eta(\nu;\vp) &= \frac{\pi \vp^2}{ \epsilon_\vp} + \frac{\eta}{N} g_\eta(\nu;\vp) 
\end{eqnarray}
All the eigenvalues $\eta$ we obtained numerically were complex; most had a negative real part which leads to exponential decay. We typically found only a handful of eigenvalues with a positive real part, and it is these that lead to exponential growth of chaotic correlations. An inverse Laplace transform $\CC(t) = \int_\nu e^{-i\nu t} \CC(\nu)$ then gives
\begin{eqnarray}
\CC(t) \sim e^{\eta t T/N }
\end{eqnarray}
which allows us to identify the eigenvalue $\eta$ as the dimensionless function $n(u,T)$ introduced earlier in Eq.(\ref{eq:lambdaL}).  

We remark that while maximal chaos is often associated with 
the absence of ``quasiparticles"\cite{SachdevBook,HartnollHolographic2016} (so presumably large scattering rate $\Gamma_p$), in equation~\ref{eqn:g-integral-eqn}
the scattering rate appears to decrease the chaos exponent in an apparent contradiction. There is however no contradiction as $\CR_1(\vl,\vp)$ which is present in Eq.\ref{eqn:g-integral-eqn} and contributes to chaos also contributes to $\Gamma_\vp$ (see Eq.\ref{eqn:scattering-rate}).

Despite its $z=2$ scaling, the quantum Lifshitz critical point has apriori a velocity scale $\sqrt{\kappa \Lambda}$. 
This is consistent with Lieb-Robinson type bounds that show the existence of a light cone even in non-relativistic systems.\cite{Bravyi2006} 
This is explicitly a UV sensitive quantity. As we argued earlier a ``thermal velocity", $v = \sqrt{r(T)} \sim \sqrt{\alpha(u,T) T}$, is induced in our model close to the QCP.

There is growing evidence that OTOCs like ours propagate ballistically in systems that exhibit a non-zero chaos exponent.\cite{Roberts2015, Shenker2014}.
Roberts and Swingle\cite{RobertsSwingle2016} have argued that the associated velocity is an effective Lieb-Robinson velocity in the IR, 
and is UV insensitive (although this depends on the considered operators, see supplemental material of Ref~\cite{RobertsSwingle2016}). 
Multiple calculations support this conjecture\cite{PatelSachdev}.
%NOTE: Roberts, Stanford and Susskind seem to define an operator spreading radius that only makes sense AFTER the scrambling time t_* ~ \beta \logN^2 has passed.

In the qLM we find numerical evidence that $C(t,\vk)$ grows exponentially with an exponent \begin{equation}
\lambda(\vk) = \lambda_L - D_L \vk^2 + \cdots
\end{equation}
 where $\lambda_L$ denotes the Lyapunov exponent.
With some additional assumptions, that will be spelled out later, it follows that 
\begin{eqnarray}
	\label{eqn:c-gaussian-form}
	C(t,\vx) &\sim \int_\vk e^{i\vk \cdot \vx} e^{\left (\lambda_L - D_L \vk^2 \right )t} \sim e^{\lambda_L t - \frac{\vx^2}{4D_L t}}
\end{eqnarray}
 and the wavefront propagates ballistically with (by definition) Butterfly velocity $v_B = \sqrt{4 D_L \lambda_L}$.

Mirroring the discussion for $\CC(t)$, for $\vq \neq 0$ we define the auxiliary function $g(\nu,\vq;\omega,\vp)$
\begin{eqnarray}
	\CC(\nu) &= \frac{1}{N} \int\frac{d\omega}{2\pi}\int_\vp \vp \cdot (\vp-\vq) g(\nu,\vq; \omega,\vp)
\end{eqnarray}
and obtain the new Bethe-Salpeter equation for Type 1 ladders:
\begin{eqnarray}
\fl	g(\nu,\vq;\omega,\vp) = \CG_R(\nu-\omega,\vq-\vp)\CG_R(\omega,\vp)  \\
\fl \times \left [ \vp \cdot (\vp - \vq)  
	+ \frac{1}{N}\int_{\omega',\vp'} \left ( \vp \cdot \vp' \right )  (\vq-\vp) \cdot (\vq-\vp')  \CG_{W,\lambda}(\omega'-\omega,\vp'-\vp)  g(\nu,\vq;\omega',\vp') \right ] \nonumber 
\end{eqnarray}

The Type 2 rungs are included by shifting 
\begin{eqnarray}
\fl	\left ( \vp \cdot \vp' \right )&  (\vq-\vp) \cdot (\vq-\vp')  \CG_{W,\lambda}(\omega'-\omega,\vp'-\vp)\nonumber \\
\fl	 & \rightarrow  \left ( \vp \cdot \vp' \right )  (\vq-\vp) \cdot (\vq-\vp')  \CG_{W,\lambda}(\omega'-\omega,\vp'-\vp) 
	+ \CG_{\rm eff}(\nu,\vq; \omega,\omega',\vp,\vp')
\end{eqnarray} 
where 
\begin{eqnarray}
\fl	\CG_{\rm eff}(\nu,\vq; \omega,\omega',\vp,\vp') = \nonumber \\
\fl \quad \int_{\omega'',\vp''} (\vq-\vp)\cdot(\vp''-\vp) \vp \cdot (\vp''-\vp) (\vq-\vp')\cdot(\vp'-\vp'') (\vp'-\vp'')\cdot \vp'\nonumber \\ 
\fl \quad	 \times  \CG_W(\omega''-\omega,\vp''-\vp) \CG_W(\omega'-\omega'',\vp'-\vp'')  \CG_{R,\lambda}(\nu-\omega'',\vq-\vp'') 
	\CG_{R,\lambda}(\omega'',\vp'')
\end{eqnarray}
A simplified expression for $\CG_{\rm eff}(\nu,\vq; \omega,\omega',\vp,\vp')$ that is more amenable to numerical evaluation is provided in Eqn~\ref{eqn:geff-q-simplified}.

Adopting the ansatz
\begin{eqnarray}
	g(\nu,\vq;\omega,\vp) &= \frac{1}{2\epsilon_\vp} \left [ g_+(\vp) \delta(\omega-\epsilon_\vp) + g_-(\vp) \delta(\omega+\epsilon_\vp) \right ]
\end{eqnarray}
we arrive at the following system of integral equations
\begin{eqnarray}
	\label{eqn:boltzmann-form-finite-q}
\fl	\left (-i\nu + i\delta\epsilon_\vq + 2\Gamma_\vp \right )g_+(\vp) = & \frac{\pi}{\epsilon_{\vq-\vp}} \vp\cdot(\vq-\vp) \nonumber \\
	 & + \frac{1}{N}\int_{\vp'} \CR_+(\vp',\vp)g_+(\vp') + \CR_-(\vp',\vp)g_-(\vp') \\
\fl	\left (-i\nu - i\delta\epsilon_\vq + 2\Gamma_\vp \right )g_-(\vp) =& \frac{\pi}{\epsilon_{\vq-\vp}} \vp\cdot(\vq-\vp) \nonumber \\
	&+ \frac{1}{N}\int_{\vp'} \CR_-(\vp',\vp)g_+(\vp') + \CR_+(\vp',\vp)g_-(\vp')
\end{eqnarray}
The integral kernels are almost identical to the earlier versions:
\begin{eqnarray}
	\CR_\pm(\nu,\vq; \vp',\vp) &= R_{1,\pm}(\nu,\vq; \vp',\vp) + R_{2,\pm}(\nu,\vq;\vp',\vp)\\
	\CR_{1,\pm}(\nu,\vq; \vl,\vp) &=  \frac{(\vq-\vl)\cdot(\vq-\vp) \,\, \vl\cdot\vp}{4 \epsilon_\vl \epsilon_\vp} 
		 \CG_{W,\lambda}(\pm \epsilon_\vl - \epsilon_\vp, \vl-\vp)\\
	\CR_{2,\pm}(\nu,\vq;\vl,\vp) &= \frac{1}{4 \epsilon_\vl \epsilon_\vp} \CG_{\rm eff} \left (\nu,\vq; \pm \epsilon_\vl, \epsilon_\vp; \vl-\vp, \frac{\vl+\vp}{2} \right )
\end{eqnarray}

We have made the following approximations in reaching the above expressions: $\epsilon_{\vp'} \epsilon_{\vq-\vp} \approx \epsilon_{\vp'} \epsilon_\vp$, 
$\epsilon_\vp-\epsilon_{\vq-\vp} \equiv \delta\epsilon_\vq  \approx \vq \cdot v_\vp$. 
We also used the fact that the Wightman functions are even in frequency and rotationally invariant.
As noted in Ref~\cite{ChowdhurySwingle2017,Stanford2016}, Eq.\ref{eqn:boltzmann-form-finite-q} has the form of a kinetic theory Boltzmann equation.
 This analogy inspires the ansatz that ``particle" density at $-\vp$ is equals ``hole" density at $\vp$, that is $g_+(-\vp) = g_-(\vp)$. 
 This finally allows us to decouple the two equations; we arrive at the following (dropping the homogeneous term) final expression:
\begin{eqnarray}
	\label{eqn:g-integral-eqn-qx}
\fl	\left (-i\nu + i \vq \cdot v_\vp  \right )g_+(\vp) =\nonumber \\
\fl \qquad \qquad \qquad  \frac{1}{N}\int_{\vp'} \left [ \CR_+(\vp',\vp) + \CR_-(-\vp',\vp) - 2N \Gamma_{\vp} \delta^{(2)}(\vp'-\vp) \right ]g_+(\vp')
\end{eqnarray}

Our identification of the Butterfly velocity following Eqn~\ref{eqn:c-gaussian-form} holds assuming the eigenvectors $g_+(\vp)$ of the equation above are weakly $\vq$ dependent.

\section{Numerical Results}

We now describe the solution of \ref{eqn:g-integral-eqn} and \ref{eqn:g-integral-eqn-qx}. It is achieved by discretizing the momentum integral on the right hand side into a sum. This turns the integral kernel into a matrix and the function $g(\nu,\vq; \vp')$ into a vector. We diagonalize this matrix. Using eigenfunctions we simplify to the form \ref{eqn:g-eigenfunction} (or similar for $\vq \neq 0$). Finally an inverse Laplace transform takes us to real time.

The numerical calculations were carried out as described below for $\alpha = 10^{-1}, 10^{-2}, 10^{-5}$: 
\begin{itemize}
	\item	We were able to eliminate the Dirac delta functions present in $\Im \, \Pi_{\lambda,R}(\nu,\vq)$ by solving the quartic polynomial in the argument. This produced four integrals over two-dimensional momentum that nevertheless still vanish in some regions.
	\item	We computed $\Im \, \Pi_{\lambda,R}(\nu,\vq)$ at fixed $\vq$ for many values of $\nu$, particularly densely around $\nu_*(\vq)=\sqrt{\alpha \vq^2 + \kappa \vq^4}$.
	\item	We rescaled the $\nu$ coordinate of these slices by $\nu_*(\vq)$ and interpolated linearly between them (in the $\vq$ direction). 
	\item	We performed the Kramers-Kronig transform as outlined in \ref{app:lambda-self-energy}.
	\item	We computed $A_\lambda$ from $\CG_{R,\lambda}$ and from this obtained the $\CR_1$ functions.
	\item	To obtain $\CR_2$ we performed yet another momentum integral with the same delta function solution as earlier. This function depends on four independent real parameters (frequency, two momentum magnitudes, a relative angle) and therefore cannot be precomputed and stored for later fast lookup.
	\item	For the $\vq=0$ case we discretized the integral equation of Eq.\ref{eqn:g-integral-eqn} and turned it into an eigenvalue equation on an adaptively spaced grid of at least 60 x 60, or as fine as necessary to achieve grid spacing independence of the result.
	\item	For $\vq \neq 0$ due to the lack of rotational invariance of Eq. \ref{eqn:g-integral-eqn-qx} we are forced to use the separate components of the momentum, and which lead to much smaller grid sizes around 25 x 25 (still leading to matrices of roughly 8000 x 8000). 
\end{itemize}

The total duration of our calculations was several thousand CPU hours. All our computations were performed with $\kappa=1$. Furthermore, we chose to study OTO correlations only of current operators, $\grad \phi$. The main result of this paper is the existence of a finite chaotic Lyapunov exponent for such operators; we express it in the form $\lambda_L = n(u,T) \frac{T}{N}$, where $n(u,T)$ is a dimensionless function that is either weakly dependent on or completely independent of temperature. We estimate an error margin of 50\% on our results for $n(u,T)$. 
Recall that the ratio $\alpha = r_{\rm eff}(u,T)/T$ is uniquely determined by $T,\kappa,u,\Lambda$. However, in light of its very weak dependence on temperature, we consider it an effective independent parameter in wide temperature ranges.

\begin{figure}[hbt]
\begin{center}
\includegraphics[scale=0.5]{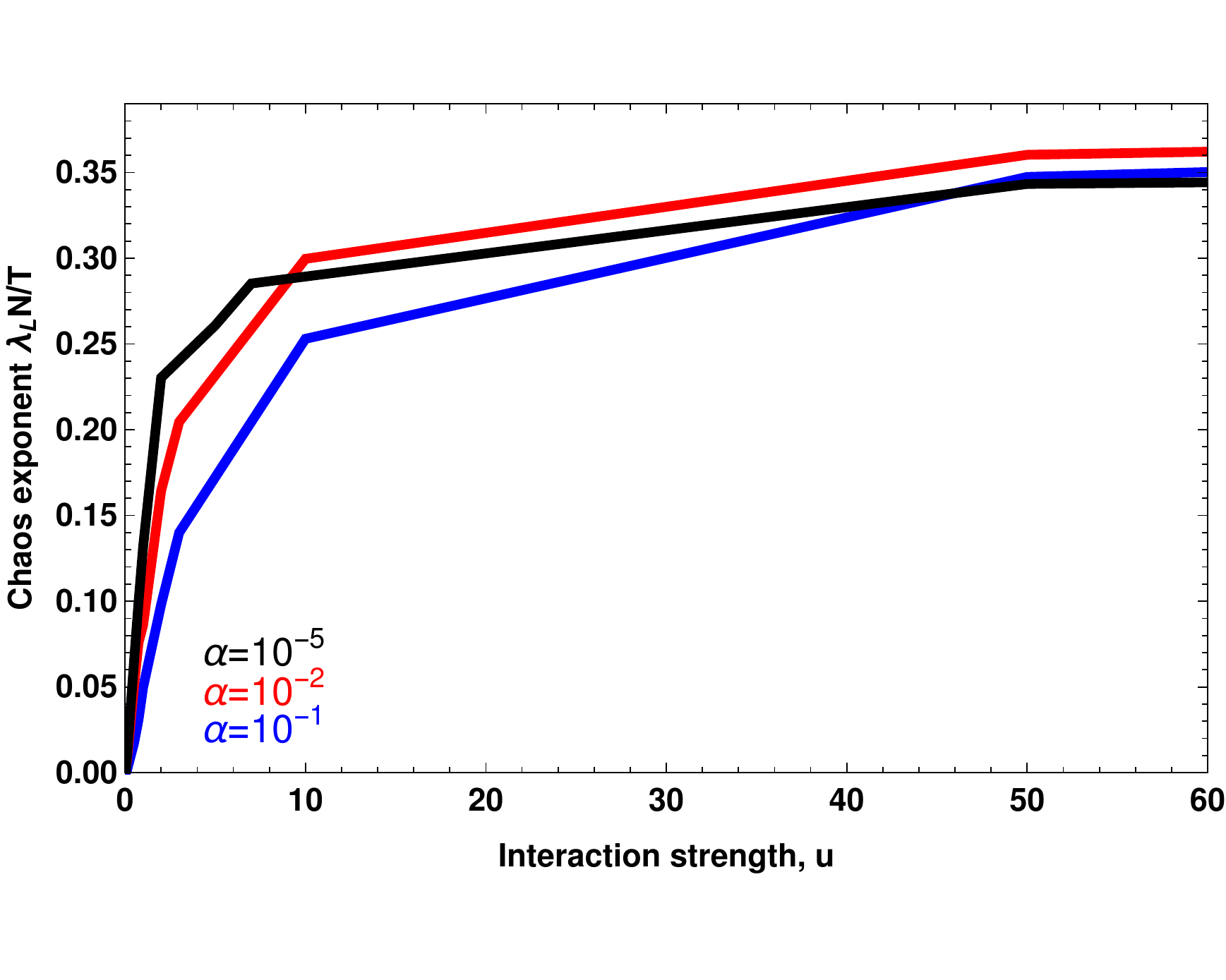}
\end{center}
\caption{Lyapunov exponent at different temperatures as a function of interaction strength.}
\label{fig:lambda-vs-int}
\end{figure}
In the {\it low temperature regime} our data (Fig.\ref{fig:lambda-vs-int}) suggests that $\lambda_L$ is weakly dependent on $\alpha$ as it goes to zero, which corresponds to decreasing temperature. At large values of the interaction strength there does not appear to be a monotonic dependence of the Lyapunov exponent on $\alpha$; our numerical error estimates preclude us from resolving this.

This same figure can be interpreted instead in the {\it weak interaction regime} when $\alpha \sim u \log(u) $ is temperature independent.
\begin{figure}[hbt]
\begin{center}
\includegraphics[scale=0.5]{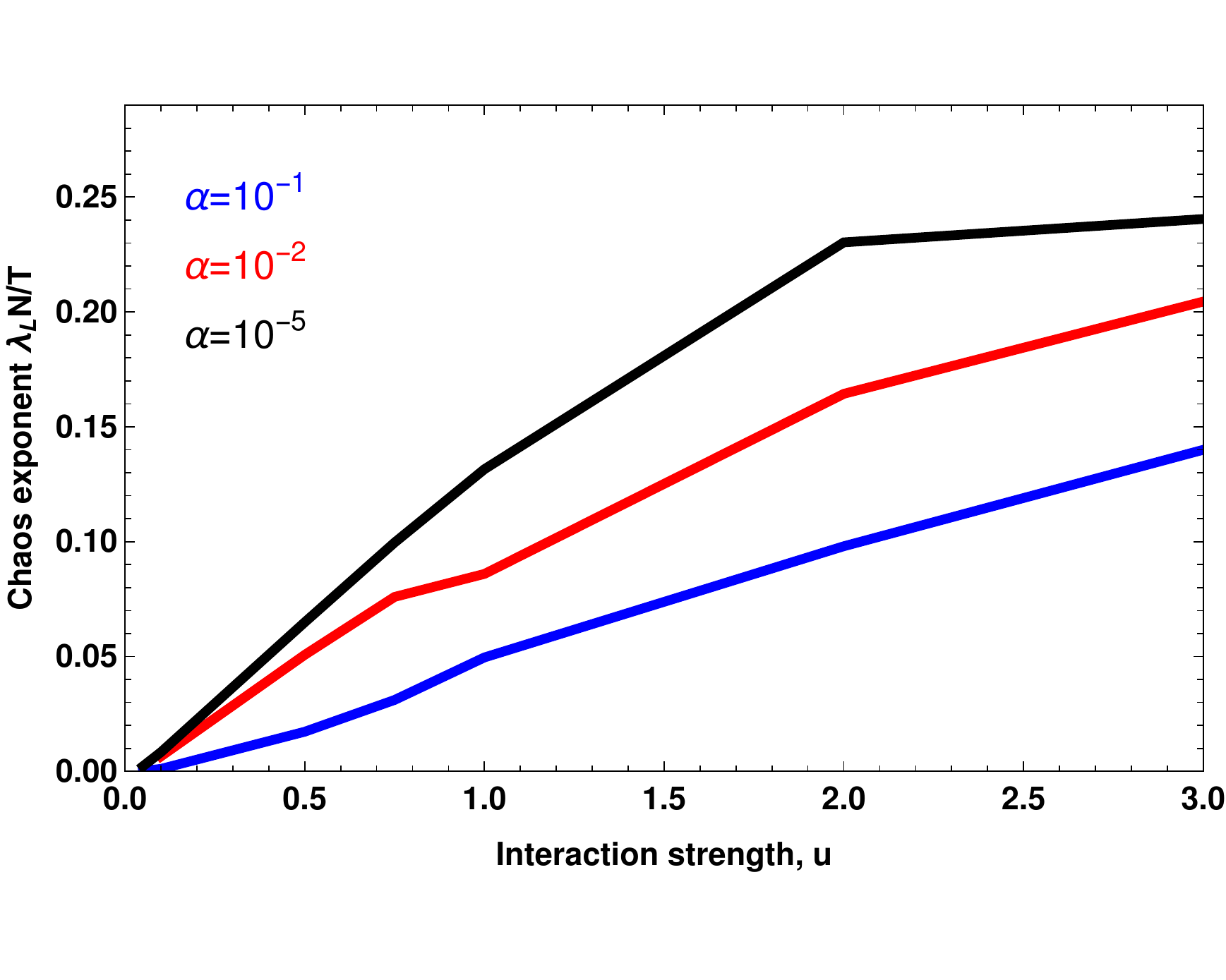}
\end{center}
\caption{Lyapunov exponent at different temperatures at small interaction strength. The non-monotonicities are due to the inaccuracy of our numerical algorithm.}
\end{figure}
As expected chaos vanishes as $u \rightarrow 0$. The saturation of the chaotic exponent at large interaction strenghts can be understood from the dressed propagator of the auxiliary field $\lambda$, (suppressing indices) of the form
\begin{eqnarray}
	\CG_\lambda &= \frac{1}{\frac{-1}{2u} - \Pi}
\end{eqnarray} 

This is the only place in the calculation where the interaction strength enters explicitly in this regime (the self-energy $\Pi$ is interaction independent). The saturation of the chaos exponent then follows from the fact that fluctuations, captured by $\Pi$, dominate the bare propagator.
An intuitive way to understand the saturation is to realize that as $u \rightarrow \infty$, field configurations with non-zero values of the current $\grad \phi$ become energetically suppressed and therefore the dynamics become more constrained.

\begin{figure}[hbt]
\begin{center}
\includegraphics[scale=0.5]{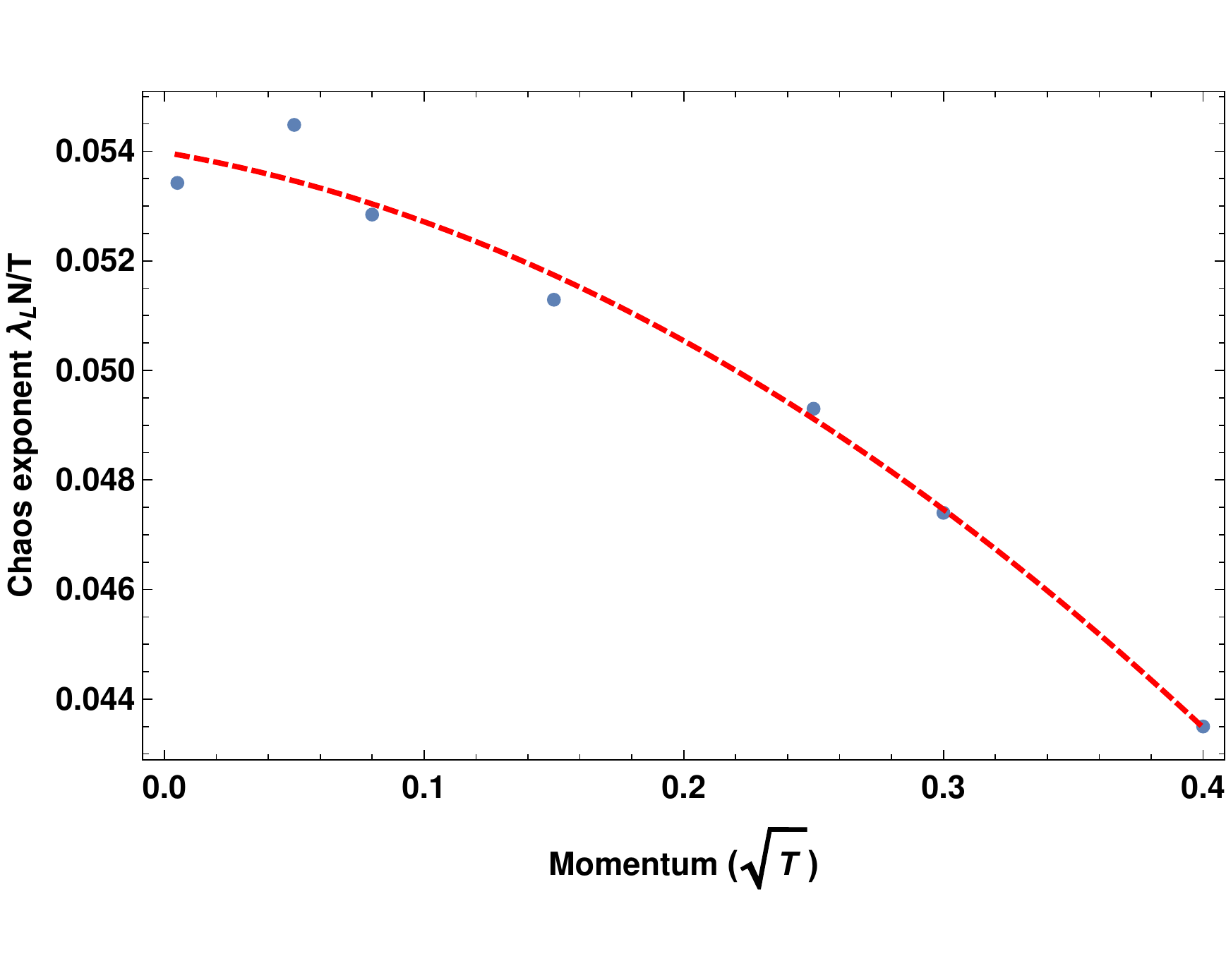}
\caption{Lyapunov exponent $\lambda(\vk)$ vs momentum, in $C(t,\vk) \sim \int d\vk \exp \left (\lambda(\vk) t \right )$, at $\alpha=0.1$ and $u=1$. The fit gives $\lambda(\vk) = 0.054 -0.008 k -0.045 k^2$, from which it follows that $v_B = 2 \sqrt{0.054 \times 0.045}\sqrt{T} = 0.1\sqrt{T}$.}
\end{center}
\end{figure}

The two independent ways of computing $\lambda_L$ (with $\vq=0$ and $\vq \geq 0$) were within 50\% of each other. Based on this we estimate our Butterfly velocities to have an error margin of about 50\%.
It is worth noting that the ``thermal velocity" $\sqrt{r(T)} = \sqrt{\alpha(T) T}$ equals $\sqrt{0.1 T} \approx 0.3 \sqrt{T}$, which is larger than largest value of the Butterfly velocity of $0.23 \sqrt{T}$ we obtained at $\alpha=1, u = 2$. We did not compute the Butterfly velocity for values other than $\alpha = 0.1$ because the computations required more accuracy for small values of $\alpha$ and we estimated their running time to be impractically long.

\begin{figure}[hbt]
\begin{center}
\includegraphics[scale=0.5]{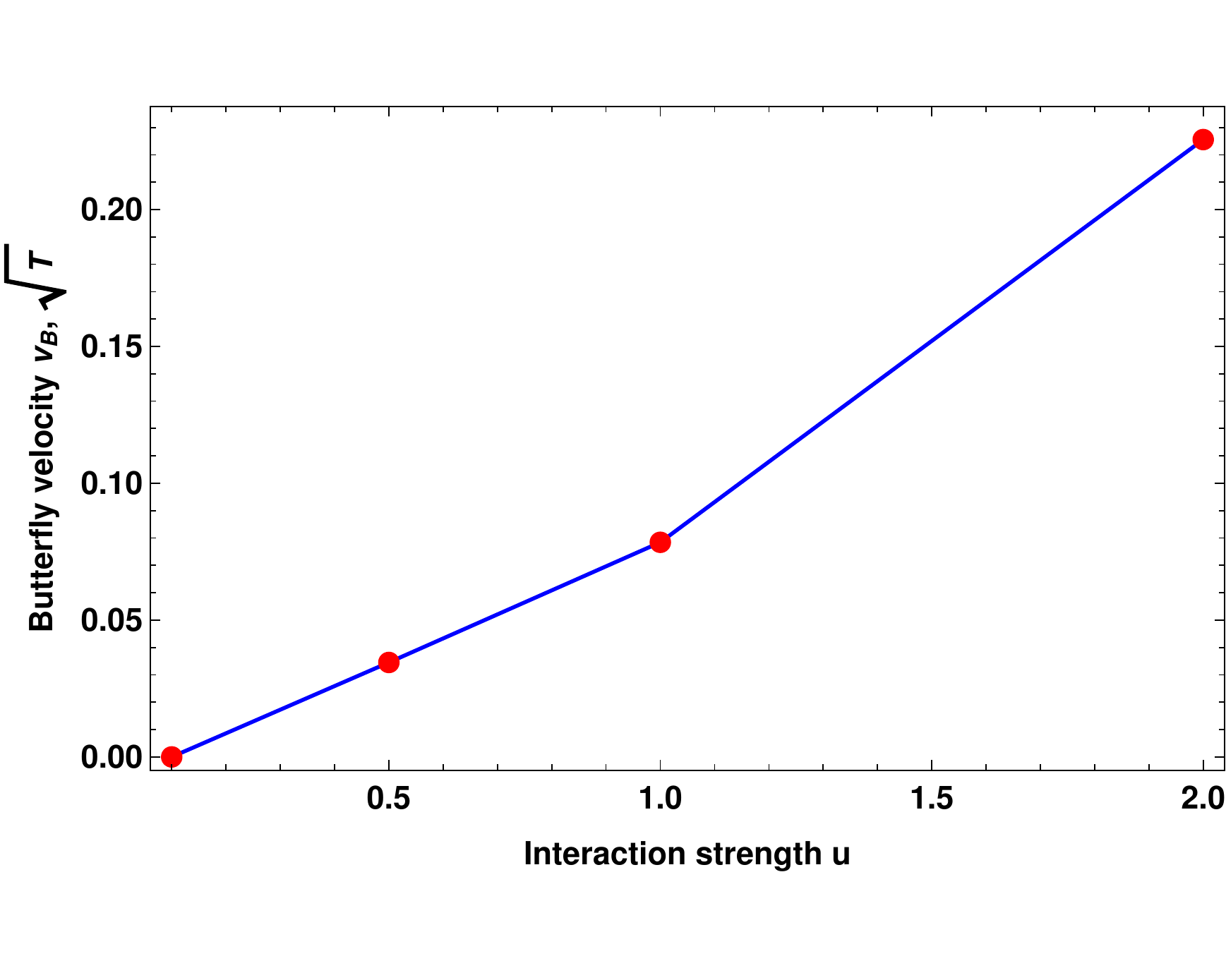}
\caption{Butterfly velocity as a function of interaction strength at $\alpha=0.1$.}
\end{center}
\label{fig:butterfly-vs-interaction}
\end{figure}

\section{Conclusions}

We performed a perturbative calculation of a out-of-time-order correlator of current operators in the non-compact quantum Lifshitz model. We extracted the numerical values of the Lyapunov exponent and Butterfly velocity for a wide range of temperatures and/or interaction strengths. 
Our results indicate that the $(\grad \phi)^4$ term is sufficient to generate chaos and is its dominant cause at sufficiently low energies. We observe that the Lyapunov exponent has a monotonic dependence on the interaction strength. In the {\it small temperature regime} it saturates at large values of interaction (which correspons to the non-linear sigma model version of the quantum Lifshitz). In the {\it weak interaction regime} the Lyapunov exponent vanishes as the interaction strength vanishes.
It is expected that for sufficiently generic interactions the Lyapunov exponent does not depend on the choice of operators in the OTO correlator. It would be interesting to verify this for our model, and for vertex operators in the compact version of qLM. 

\ack
%\begin{acknowledgments}
We thank Yuxuan Wang, Hart Goldman, Cristian Gaidau, Thomas Scaffidi, and Juan Maldacena for discussions. We also thank Debanjan Chowdhury and Brian Swingle for making available their MATLAB code for the O(N) model \cite{ChowdhurySwingle2017}. This work was supported in part by the Gordon and Betty Moore Foundation EPiQS Initiative through Grant No. GBMF4305 (EP) and by the National Science Foundation through the grant DMR 1725401 (EF).
%\end{acknowledgments}

\appendix

\section{Saddle-point solution at $N=\infty$}

\label{app:saddle-point-solution}

We rewrite Eq.\ref{eqn:phi-lambda-N} in a form convenient for integrating out $\phi$
\begin{eqnarray}
\fl	S_{\phi,\lambda} = \frac{1}{2}\int d\tau_1  d^2\vr_1 d\tau_2 d^2\vr_2  & \phi(\tau_1,x_1)G^{-1}(\tau_1,\vr_1;\tau_2,\vr_2)\phi(\tau_2,x_2)	\nonumber \\
		&+\int d\tau d^2\vr \frac{\lambda^2(\tau,x)}{4u}
\end{eqnarray}
where now
\begin{eqnarray}
\fl	G^{-1}&(\tau_1,x_1;\tau_2,x_2) = \delta(\tau_1-\tau_2)\delta^{(2)}(\vr_1 -\vr_2) \nonumber \\
\fl & \times \left ( -\partial_{\tau_1}^2  - r_{qc} \grad^2_{\vr_1} + \kappa \grad^4_{\vr_1} - \frac{1}{\sqrt{N}} \grad \lambda(\tau_1,\vr_1) \cdot \grad - \frac{1}{\sqrt{N}}\lambda(\tau_1,\vr_1)\grad^2 \right )
\end{eqnarray}

From now on we will write ${\bf i}$ for the vector $(\tau_i, \vr_i)$ in order to make things more readable.

We divide the $\phi_a$ into $N-1$ components $\phi_\perp$ and $\pi$ and we choose to integrate out $\phi_\perp$ and subsequently use the $\pi\pi$ correlations as a proxy for the $\phi\phi$ correlations of the original theory. At large $N$ the difference between $N$ and $N-1$ is unimportant.

\begin{eqnarray}
\label{eqn:effective-pi-lambda}
\fl	S_{\rm eff}(\pi,\lambda) &=  \int d\tau d^2\vr  \frac{1}{2}  \left [\left (\partial_\tau \pi \right )^2  + r_{qc}\left ( \grad \pi\right )^2 +   \kappa \left ( \grad^2 \pi \right )^2  + \frac{\lambda(\tau,\vr)}{\sqrt{N}} (\grad\pi)^2 \right ] \nonumber \\
\fl &- \int d\tau d^2\vr \frac{\lambda^2(\tau,x)}{4u} + \frac{N-1}{2}\ln \det G^{-1}
\end{eqnarray}

The saddle-point equation for a spatially uniform $\lambda$ is easiest found assuming all components of $\phi_a$ have been integrated out
\begin{eqnarray}
\fl	\frac{\delta S_{\rm eff}}{\delta \lambda} &= 0 \\
\fl	& = -\beta L^2 \frac{2 \lambda}{4u} + \frac{N}{2} \frac{\delta \ln\det G^{-1}}{\delta \lambda} \\
\fl	&=  -\beta L^2 \frac{2 \lambda}{4u} +  \frac{\sqrt{N}}{2} \beta L^2 \frac{1}{\beta} \sum_{\omega_n} \int \frac{d^2\vq}{(2\pi)^2} \frac{\vq^2}{\omega_n^2 + (r_{\rm qc} +\lambda/\sqrt{N} )\vq^2 + \kappa \vq^4} 
\end{eqnarray}
Hence,
\begin{equation}
\fl	\lambda  = {u \sqrt{N}} \measure \frac{1}{\omega_n^2 + (r_{\rm qc} +\lambda/\sqrt{N} )\vq^2 + \kappa \vq^4} 
\end{equation}

It will be convenient to define $r_{\rm eff} = r_{qc} + \lambda/\sqrt{N}$ and replace $\lambda$ in terms of it:
\begin{eqnarray}
	r_{\rm eff}-r_{qc} &= u \measure \frac{\vq^2}{\omega_n^2 + r_{\rm eff}^2 \vq^2 + \kappa \vq^4} 
\end{eqnarray}
Using the identity
\begin{eqnarray}
	\frac{1}{\beta}\sum_{\omega_n} \frac{1}{\omega_n^2 + a^2} &= \frac{1}{a} \left [ \frac{1}{2} + \frac{1}{e^{\beta a} - 1}\right ]
\end{eqnarray}
we arrive at the final form of this self-consistent equation
\begin{eqnarray}
\label{eqn:saddle-point-eqn}
\fl	r_{\rm eff}-r_{qc} &= u \int \frac{d^2 \vq}{(2\pi)^2} \frac{\vq^2}{\sqrt{r_{\rm eff}\vq^2 + \kappa \vq^4}} \left [ \frac{1}{2} + \frac{1}{\exp\left (\beta \sqrt{r_{\rm eff}\vq^2 + \kappa \vq^4}\right )-1}\right ]
\end{eqnarray}

The physics is now most transparent. $r_{\rm eff}$ is the effective stifness/velocity of the theory, and must vanish at the zero-temperature QCP. This implies $r_{qc}$ must have the value 
\begin{eqnarray}
	-r_{qc} &= u \measure \frac{\vq^2}{\omega_n^2 + \kappa \vq^4} = \frac{u}{8} \int_\Lambda\frac{d^2\vq}{(2\pi)^2} \frac{\vq^2}{\sqrt{\kappa \vq^4}}
\end{eqnarray} 
which needs to be regularized with a UV cutoff.

It is difficult to solve Eq.\ref{eqn:saddle-point-eqn} directly, but it can be guessed that $r_{\rm eff}$ vanishes faster than $T$ due to log corrections, and therefore in the regime where $r_{\rm eff} \ll \rk T$ the equation becomes
\begin{equation}
\label{app:reff-equation}
r_{\rm eff} = \frac{uT}{4\pi \kappa} \frac{\ln\left ( \frac{\rk T}{r_{\rm eff}}\right )}{1+\frac{u}{16\pi\kappa^{3/2}}\left [ \ln\left (\frac{4 \kappa \Lambda^2}{r_{\rm eff}}\right )\right ]}
\end{equation}

When the log in the denominator dominates, and after rescaling $\bar{r} = r_{\rm eff}/\rk T$, the equation simplifies to
\begin{equation}
\rbar = 4 \frac{\ln\left ( \frac{1}{\rbar}\right )}{\ln\left (\frac{4 \Lambda^2}{\rbar \rk T}\right )} = 4 \frac{\ln\left ( \frac{1}{\rbar}\right )}{\ln\left (\frac{4 \Lambda^2}{\rk T}\right ) - \ln \rbar } 
\end{equation}

The condition that in Eqn~\ref{app:reff-equation} the denominator is dominated by the log implies in particular that $\rbar \ll \Lambda^2/\rk T$, which implies
\begin{equation}
\rbar \approx 4 \frac{\ln\left ( \frac{1}{\rbar}\right )}{\ln\left (\frac{4 \Lambda^2}{\rk T}\right )  } 
\end{equation}

Ignoring the log in the numerator we see that a putative solution to the equation is $\rbar = 4/\ln\left (\frac{4 \Lambda^2}{\rk T}\right )$. We substitute it into the original equation and find that it is wrong by a prefactor $ \ln \ln\left (\frac{4 \Lambda^2}{\rk T}\right ) $ which we incorporate into the new trial solution $\rbar$. Another substitution shows that the solution is now correct up to factors of $\left [ \ln \ln\left (\frac{4 \Lambda^2}{\rk T}\right ) \right ]^2/\ln\left (\frac{4 \Lambda^2}{\rk T}\right )^2 \approx \rbar^2 $ which is guaranteed to be small because $\rbar \ll 1$.

\section{1/N corrections}
\label{app:1-n-corrections}

Rewrite Eq.\ref{eqn:effective-pi-lambda} with the saddle-point solution substituted and simultaneously denote deviations from it by $\lambda$: 
\begin{eqnarray}
\label{eqn:effective-pi-saddle-lambda}
\fl	\frac{\lambda_{\rm sp}}{\sqrt{N}} &= r_{\rm eff} - r_{qc} \nonumber \\
\fl	S_{\rm eff}(\pi,\lambda_{\rm sp}+\lambda) &=  \int d\tau d^2\vr  \frac{1}{2}  \left [\left (\partial_\tau \pi \right )^2  + r_{\rm eff}\left ( \grad \pi\right )^2 +   \kappa \left ( \grad^2 \pi \right )^2  + \frac{\lambda(\tau,\vr)}{\sqrt{N}} (\grad\pi)^2 \right ] \nonumber \\
	&- \int d\tau d^2\vr \frac{\left (\lambda_{\rm sp} + \lambda\right )^2(\tau,x)}{4u}  + \frac{N-1}{2}\ln \det G^{-1} \nonumber
\end{eqnarray}
where now
\begin{eqnarray}
\fl	G^{-1}(1,2) &= \delta(1-2)\left ( -\partial_{1}^2  - r_{\rm eff} \grad^2_{1} + \kappa \grad^4_{1} - \frac{1}{\sqrt{N}} \grad \lambda(1) \cdot \grad_1 - \frac{1}{\sqrt{N}}\lambda(1)\grad_1^2 \right )
\end{eqnarray}

Now expand $\ln \det G^{-1}$ to second order in $\lambda(\tau,\vx)$ around the saddle-point value.
\begin{eqnarray}
	\ln\det G^{-1} &= \tr \ln G^{-1} \\
	&= \tr\ln G_0^{-1}\left (1+G_0 B\right )  = \tr\ln G_0^{-1}+\tr\ln\left (1+G_0 B\right ) \\
	&=  \tr\ln G_0^{-1} + \tr G_0 B - \frac{1}{2} \tr (G_0 B)^2 + \cdots
\end{eqnarray}
The first order term will vanish when combined with the variation of the rest of the action due to the stationary phase condition. Finally, the second order term gives
\begin{eqnarray}
	G_0^{-1}(1,2) &= \delta(1-2)\left ( -\partial_{1}^2  - r_{\rm eff} \grad^2_{1} + \kappa \grad^4_{1} \right ) \\
	B(1,2) &= \frac{-1}{\sqrt{N}} \delta(1-2) \left [ \grad \lambda(1) \cdot \grad + \lambda(1) \grad^2 \right ] \\
	\tr (G_0 B)^2  &= \frac{1}{N} \int d1\,d2\,d3\,d4\, G_0(1,2) B(2,3) G_0(3,4) B(4,1) \\
	&= \frac{1}{N}\int d1\,d2\, \lambda(1) \lambda(2) \grad_1 \grad_2 G_0(1,2)  \grad_1 \grad_2 G_0(2,1)
\end{eqnarray}
with which the effective action for $\lambda$ at order $O(N^0)$ becomes
\begin{eqnarray}
\fl	S_{\rm eff}(\pi,\lambda) &=  \int d\tau d^2\vr  \frac{1}{2}  \left [\left (\partial_\tau \pi \right )^2  + r_{\rm eff}\left ( \grad \pi\right )^2 +   \kappa \left ( \grad^2 \pi \right )^2  + \frac{\lambda(\tau,\vr)}{\sqrt{N}} (\grad\pi)^2 \right ]  \nonumber \\
\fl	&- \measure \lambda(\omega_n,\vq) \lambda(-\omega_n,-\vq)  G_1^{-1}(i\omega_n,\vq) \nonumber \\
\fl G_1^{-1}(i\omega_n,\vq) &= \left [ -\frac{1}{4u}+ \meas{\omega'}{p} \left [ (\vp+\vq) \cdot \vp \right ]^2 G_0(\omega'_n+\omega_n,\vp+\vq)G_0(\omega'_n,\vp) \right ] \nonumber
\end{eqnarray}

\section{Derivation of OTO rules}
\label{app:oto-rules}

We work in the interacting picture, where the time-evolution operator is 
\begin{eqnarray}
	U_I &= \mathcal{T} \exp \left (  \frac{i}{2 \sqrt{N}} \sum_a \int_0^t ds \int_\vx \lambda_0(\vx,s) \left ( \grad \phi_a(\vx,s \right )^2 \right )
\end{eqnarray}
Its expansion looks like
\begin{eqnarray}
\fl	U_I = 1 + \frac{i}{2\sqrt{N}}&\sum_a \int_{s_1}  \int_\vx \lambda_0(\vx,s) \left ( \grad \phi_a(\vx,s \right )^2 & \nonumber \\
\fl	&+ \left (\frac{i}{2\sqrt{N}} \right )^2 \sum_{a,b}  \int^t ds_1 \int^{s_1} ds_2 \int_{\vy_1, \vy_2}  \lambda_0(\vy_1,s_1) \left ( \grad \phi_a(\vy_1,s_1 \right )^2 \nonumber \\ 
 & \qquad \qquad \qquad \times \lambda_0(\vy_2,s_2) \left ( \grad \phi_b(\vy_2,s_2 \right )^2 
\end{eqnarray}
We will need an expansion of $U_I^\dagger \phi_0(\vx,t) U_I$, for which it will be useful to first establish a few identities:
\begin{eqnarray}
	\left [ \lambda_0, \phi_0 \right ] &= 0 \\
	\left [ \phi_{0,a}(\vx,t), \phi_{0,b}(0) \right ] &= i \CG_R^{ab} (\vx,t)\\
	\left [ \phi_{0,a}(\vx,t), \grad \phi_{0,b}(\vy) \right ] &=  \grad_y i \CG_R^{ab} (\vx-\vy,t)\\ 
	\left [ \grad \phi_{0,a}(\vx,t), \grad \phi_{0,b}(\vy) \right ] &=  \grad_x \grad_y i\CG_R^{ab} (\vx-\vy,t)\\	
	\left [ \phi_{0,a}(\vx,t), \left ( \grad \phi_{0,b}(\vy) \right )^2 \right ] &= 2\grad_y i \CG_R^{ab} (\vx-\vy,t) \cdot \grad_y \phi_{0,b}(\vy) \\
	\left [ \grad \phi_{0,a}(\vx,t), \left ( \grad \phi_{0,b}(\vy) \right )^2 \right ] &= 2\grad_x \grad_y i \CG_R^{ab} (\vx-\vy,t) \cdot \grad_y \phi_{0,b}(\vy) \\
\end{eqnarray}

Finally, we got back to $U_I^\dagger \phi_0(\vx,t) U_I$:
\begin{eqnarray*}
\fl	U_I^\dagger \phi_0(\vx,t) U_I&= \phi_0(\vx,t) + \frac{i}{2\sqrt{N}}\sum_a \int^t ds \int_\vy \left [ \grad \phi_0(\vx,t), \lambda_0(\vy,s) \left ( \grad \phi_{0,a}(\vy,s) \right )^2 \right ] + \cdots
\end{eqnarray*}
Which means the first-order term in $C(t,\vx)$ is
\begin{eqnarray*}
\fl	C_1(t,\vx_1-\vx_2) &= \frac{-1}{N^2} \left ( \frac{i}{2\sqrt{N}} \right )^2 \sum_{a,b,c,d} \int_{s_1,s_2} \int_{\vx_1,\vx_2} \\
	& {\rm Tr} \lbrace \sqrt{\rho} \left [ \left [ \grad \phi_a(\vx_1,t), \lambda_0(\vy_1,s_1) \left ( \grad \phi_{0,c}(\vy_1,s_1) \right )^2 \right ], \grad \phi_b(\vx_2) \right ]  \\
	& \sqrt{\rho} \left [ \left [ \grad \phi_a(\vx_1,t), \lambda_0(\vy_2,s_2) \left ( \grad \phi_{0,c}(\vy_2,s_2) \right )^2 \right ], \grad \phi_b(\vx_2) \right ] \rbrace \\
	&= \frac{-1}{N^2} \left ( \frac{i}{2\sqrt{N}} \right )^2 \sum_{a,b,c,d} \int_{s_1,s_2} \int_{\vx_1,\vx_2} \\
	&{\rm Tr} \lbrace \rho  \lambda_0(\vy_1,s_1) \lambda_0(\vy_2,s_2) \rbrace  \, 2\grad_x \grad_y i \CG_R^{ac} (\vx_1-\vy_1,t-s_1) \cdot \grad_y \grad_x i\CG^{cb}(\vy_1-\vx_2,s_1) \\
	& \, 2\grad_x \grad_y i \CG_R^{ad} (\vx_1-\vy_2,t-s_2) \cdot \grad_y \grad_x i\CG^{db}(\vy_2-\vx_2,s_2) \\
	&= \frac{-1}{N^2} \left ( \frac{i}{2\sqrt{N}} \right )^2 \sum_{a,b,c,d} \int_{s_1,s_2} \int_{\vx_1,\vx_2} \CG^W_\lambda(\vy_1-\vy_2, s_1-s_2)\\
	& \, 2\grad_x \grad_y i \CG_R^{ac} (\vx_1-\vy_1,t-s_1) \cdot \grad_y \grad_x i\CG^{cb}(\vy_1-\vx_2,s_1) \\
	& \, 2\grad_x \grad_y i \CG_R^{ad} (\vx_1-\vy_2,t-s_2) \cdot \grad_y \grad_x i\CG^{db}(\vy_2-\vx_2,s_2) 	
\end{eqnarray*}
Note that above the $\grad_x$ and $\grad_y$ are dotted with themselves. Another observation is that
\begin{eqnarray}
	\frac{1}{N^3}\sum_{abcd} \delta_{ac} \delta_{cb} \delta_{ad} \delta_{db} = \frac{1}{N^3}\sum_{ab} \delta_{ab} = \frac{1}{N^2}
\end{eqnarray}

Finally, we Fourier transform and obtain:
\begin{eqnarray}
\fl	C_1(\nu) &= \int_t \int_{\vx_1-\vx_2} C_1(t,\vx_1-\vx_2) \nonumber \\
\fl	&= \frac{1}{N^2} \int_{\vp,\vp'} \left ( \vp \cdot \vp'\right )^4 \int_{\omega,\omega'} \CG_R(\vp,\omega)\CG_R(-\vp,\nu-\omega)  \nonumber \\
\fl & \qquad \qquad \times \CG^W_\lambda(\vp-\vp',\omega-\omega') \CG_R(\vp',\omega')\CG_R(-\vp',\nu-\omega')
\end{eqnarray}
Had we chosen to instead dot product the gradient terms of $\grad_{x_1}$ (and $\grad_{x_2}$) with $\grad_{x_1}$ (and $\grad_{x_2}$) from other commutator, we would have had a different momentum-dependent numerator: $ \left ( \vp \cdot \vp' \right )^2 \vp^2 \vp'^2$:
\begin{eqnarray}
\fl	C_1(\nu) &= \frac{1}{N^2} \int_{\vp,\vp'} \vp^2 \left ( \vp \cdot \vp'\right )^2 \vp'^2 \int_{\omega,\omega'} \CG_R(\vp,\omega)\CG_R(-\vp,\nu-\omega) \CG^W_\lambda(\vp-\vp',\omega-\omega') \nonumber \\
\fl & \qquad \qquad \times \CG_R(\vp',\omega')\CG_R(-\vp',\nu-\omega')
\end{eqnarray}
Or generalized:
\begin{eqnarray}
\fl	C_{1(ijij)}(\nu,\vq) &= \frac{1}{N^2} \int_{\vp,\vp'} \vp \cdot (\vq-\vp) (\vq-\vp)\cdot(\vq-\vp') (\vq-\vp')\cdot \vp' \vp \cdot \vp' \\
\fl	& \times \int_{\omega,\omega'} \CG_R(\vp,\omega)\CG_R(-\vp,\nu-\omega) \CG^W_\lambda(\vp-\vp',\omega-\omega') \nonumber \\
 & \qquad \qquad \times \CG_R(\vp',\omega')\CG_R(-\vp',\nu-\omega')
\end{eqnarray}

This gives a diagram with one rung of Type 1. We can read off the Feynman rules from the above. They are:
\begin{itemize}
	\item	Draw a diagram
	\item	Each line must be directed: left to right, and top to down. This determines whether a momentum/frequency is incomming or outgoing. The sum of incomming must equal sum of outgoing.
	\item	Assign a value of $\frac{i \vk_1 \cdot \vk_2}{\sqrt{N}} \delta_{a,b}$ to each vertex, where the momenta belong to $\phi$.
	\item	Multiply overall diagram by $ \left (\vk_i \cdot \vk_f \right )^2$ for $C_{iijj}$ or $\vk_i^2 \vk_f^2$ for $C_{ijij}$, where the momenta $\vk_i,\vk_f$ emmanate from $\vx_1$ and $\vx_2$ correspondingly.
	\item	Retarded propagators for horizontal lines, Wightman for vertical.
\end{itemize}

To verify the rules for diagrams with Type 2 rungs we must expand $U_I^\dagger \phi U_I$ to second order.
\begin{eqnarray*}
\fl	\left [ U_I^\dagger \phi_0(x,t) U_I \right ]_2 &= \left (\frac{i}{2 \sqrt{N}} \right )^2 \sum_{bc} \int_0^t ds_1 \int_0^{s_1} ds_2 \int_{\vy_1,\vy_2} \\
	& \left [ \left [ \phi_0(\vx,t), \lambda_0(s_1,\vy_1) \left (\grad \phi_{0,b}(s_1,\vy_1) \right )^2 \right ], \lambda_0(s_2,\vy_2) \left (\grad \phi_{0,c}(s_2,\vy_2) \right )^2  \right ]
\end{eqnarray*}

The commutator in the term above becomes
\begin{eqnarray*}
\fl	[[,],] &= 2 \grad_{y_1} i\CG_R(\vx - \vy_1,t-s_1) \cdot \grad_{y_1} \\
\fl & \qquad \qquad \times \grad_{y_2} 2i\CG_R(\vy_1-\vy_2,s_1-s_2) \cdot \grad_{y_2} \phi(\vy_2,s_2) \lambda(s_1)\lambda(s_2) \\
\fl & + 2 \grad_{y_1} i\CG_R(\vx-\vy_1,t-s_1) i\CG_{R,\lambda}(\vy_1-\vy_2,s_1-s_2) \left ( \grad_y \phi(\vy_2,s_2) \right )^2 \grad_{y_1} \phi(\vy_1)	
\end{eqnarray*}
where the $\grad$'s of $\vy_1$ are contracted amonst themselves, similarly for $\vy_2$.

Finally, we must actually consider $ \left [ \left [ U_I^\dagger \phi_0(x_1,t) U_I \right ]_2, \grad_{x_2} \phi(x_2,0) \right ]$ on each time fold.
\begin{eqnarray}
	\label{eq:second-order-commutator}
\fl	\left [ [[,],], \grad_{x_2} \phi(x_2,0) \right ] &= 
	2 \grad_{y_1} i\CG_R(\vx - \vy_1,t-s_1) \cdot \grad_{y_1} \\
\fl & \qquad \times 	\grad_{y_2} 2i\CG_R(\vy_1-\vy_2,s_1-s_2) \cdot \grad_{y_2}  \nonumber \\
\fl & \qquad \qquad \times \grad_{x_2} i\CG_R(\vy_2-\vx_2,s_2) \lambda(s_1)\lambda(s_2)  \nonumber \\
\fl	&+ 2 \grad_{y_1} i\CG_R(\vx-\vy_1,t-s_1) i\CG_{R,\lambda}(\vy_1-\vy_2,s_1-s_2)  \nonumber \\
\fl & \quad \qquad	\times \left [ \left ( \grad_y \phi(\vy_2,s_2) \right )^2 \grad_{y_1} \phi(\vy_1),\grad_{x_2} \phi(x_2,0) \right ] 
\end{eqnarray}
where 
\begin{eqnarray*}
\fl \left [ \left ( \grad_y \phi(\vy_2,s_2) \right )^2 \grad_{y_1} \phi(\vy_1),\grad_{x_2} \phi(x_2,0) \right ] =
	\grad_x \grad_{y_1} i\CG_R(\vy_1-\vx_2,s_1) \left ( \grad_{y_2} \phi\right )^2 & \\
	+ 2 \grad_x \grad_{y_2} i\CG_R(\vy_2-\vx_2,s_2) \cdot \grad_{y_2} \phi(\vy_2) \grad_{y_1} \phi(\vy_1) &
\end{eqnarray*}

The first line in Eq.\ref{eq:second-order-commutator} gives the diagram with two rungs of Type 1 (if the $\lambda$'s are contracted between the time folds), or a correction to the self-energy of $\phi$. The second line can be verified to give a diagram with a single Type 2 rung, and a correction to the vertex (which is $1/N$ suppressed) in a Type 1 diagram. Furthermore, there are two types of Type 2 diagrams, ladder type, and crossed type. This latter is argued to be kinematically suppressed and is subsequently ignored.

Putting it all together, for a diagram with a single uncrossed Type 2 rung:
\begin{eqnarray*}
\fl	C_2(t,\vx_1-\vx_2) &= \frac{-1}{N^2} \left ( \frac{i}{2\sqrt{N}} \right )^4 \sum_{abcdef} \int_{s_1,s_2,s_1',s_2'}\int_{\vy_1,\vy_2,\vy_1',\vy_2'} \\
\fl	& \times 2 \grad_{x_1} \grad_{y_1} i\CG_R(\vx_1-\vy_1,t-s_1) i\CG_{R,\lambda}(\vy_1-\vy_2,s_1-s_2)  \nonumber \\ 
\fl	& \times 2 \grad_{x_2} \grad_{y_2} i\CG_R(\vy_2-\vx_2,s_2) 2 \grad_{x_1}\grad_{y_1'} i\CG_R(\vx_1-\vy_1',t-s_1') \\
\fl	& \times  i\CG_{R,\lambda}(\vy_1-\vy_2',s_1'-s_2')  2 \grad_{x_2} \grad_{y_2'} i\CG_R(\vy_2'-\vx_2,s_2') \nonumber\\
\fl & \times \grad_{y_1}\grad_{y_1'} \CG_W(\vy_1-\vy_1',s_1-s_1') \grad_{y_2}\grad_{y_2'} \CG_W(\vy_2-\vy_2',s_2-s_2')
\end{eqnarray*}

which in the end gives (with all factors of $i,2,N$ upfront):
\begin{eqnarray*}
\fl	C_2(t,\vx_1-\vx_2) &= \frac{1}{N^2} \int_{s_1,s_2,s_1',s_2'}\int_{\vy_1,\vy_2,\vy_1',\vy_2'} \\
	& \times  \grad_{x_1} \grad_{y_1} \CG_R(\vx_1-\vy_1,t-s_1) \CG_{R,\lambda}(\vy_1-\vy_2,s_1-s_2)    \nonumber \\ 
	& \times \grad_{x_2} \grad_{y_2} \CG_R(\vy_2-\vx_2,s_2) \grad_{x_1} \grad_{y_1'} \CG_R(\vx_1-\vy_1',t-s_1')   \nonumber\\
	& \times \CG_{R,\lambda}(\vy'_1-\vy_2',s_1'-s_2')  \grad_{x_2} \grad_{y_2'} \CG_R(\vy_2'-\vx_2,s_2') \\
	& \grad_{y_1}\grad_{y_1'} \CG_W(\vy_1-\vy_1',s_1-s_1') \grad_{y_2}\grad_{y_2'} \CG_W(\vy_2-\vy_2',s_2-s_2')
\end{eqnarray*}

which in momentum space becomes
\begin{eqnarray*}
\fl	C_2(\nu) &= \frac{1}{N^2} \int_{\vp,\vp',\vp''}\int_{\omega,\omega',\omega''} \left ( \vp \cdot \vp' \right )^2 \left ( \vp \cdot (\vp''-\vp) \right )^2 \left (  \vp' \cdot (\vp'-\vp'') \right )^2 \\
\fl	& \qquad \times \CG_R(\nu-\omega,-\vp) \CG_R(\omega,\vp) \CG_W(\omega''-\omega,\vp''-\vp) \\
\fl & \qquad \times \CG_{R,\lambda}(\nu-\omega'',-\vp'') \CG_{R,\lambda}(\omega'',\vp'') \CG_R(\nu-\omega',-\vp') \CG_R(\omega',\vp') 
\end{eqnarray*}

The $1/N$ counting follows from
\begin{eqnarray}
	\frac{1}{N^2} \left ( \frac{1}{\sqrt{N}}\right )^4 \sum_{abcdef} \delta_{ac} \delta_{cd} \delta_{da} \delta_{be}\delta_{ef}\delta_{fb} = \frac{1}{N^2}
\end{eqnarray}

such that we can write the diagram with one Type 2 rung as
\begin{eqnarray}
\fl	C_{2(iijj)}(\nu) &= \frac{1}{N^2} \int_{\vp,\vp'} \left ( \vp \cdot \vp' \right )^2 \int_{\omega,\omega'}  \, \CG_R(\nu-\omega,-\vp) \CG_R(\omega,\vp) \CG_{\rm eff}(\omega',\omega,\vp',\vp) \nonumber \\ 
	& \times \CG_R(\nu-\omega',-\vp') \CG_R(\omega',\vp') 
\end{eqnarray}

Compare it with
\begin{eqnarray}
\fl	C_{1(iijj)}(\nu) &= \frac{1}{N^2} \int_{\vp,\vp'} \left ( \vp \cdot \vp'\right )^4 \int_{\omega,\omega'} \CG_R(\vp,\omega)\CG_R(-\vp,\nu-\omega) \nonumber \\
 & \times \CG^W_\lambda(\vp-\vp',\omega-\omega')  \CG_R(\vp',\omega')\CG_R(-\vp',\nu-\omega')
\end{eqnarray}
we notice that replacing $\left ( \vp \cdot \vp'\right )^2 \CG^W_\lambda(\vp'-\vp,\omega'-\omega)$ in $C_1$ by $\left ( \vp \cdot \vp'\right )^2 \CG^W_\lambda(\vp'-\vp,\omega'-\omega) + \CG_{\rm eff}(\omega',\omega,\vp',\vp)$ allows us to treat both one-rung diagrams simultaneously.

Recall that in defining the OTO correlator we had a choice in how to contract the spatial derivatives:
\begin{eqnarray}
\fl	C_{ijkl}(t_1-t_2,\vx_1 - \vx_2)  = \nonumber \\
\fl  \qquad \frac{1}{N^2}\sum_{a,b}{\rm Tr} \lbrace \sqrt{\rho} \left [ \partial_i \phi_a(\vx_1,t_1), \partial_j \phi_b(\vx_2,t_2) \right ]^\dagger \sqrt{\rho} \left [ \partial_k \phi_a(\vx_1,t_1), \partial_l \phi_b(\vx_2,t_2) \right ] \rbrace 
\end{eqnarray}
{\bf N.B.} So far we have worked with the contraction $C = \sum_{i,k} C_{iikk}$, but another choice is $C = \sum_{i,j} C_{ijij}$. I will work with the latter from now on due to certain simplifications.

Using the $ijij$ contraction:
\begin{eqnarray}
\fl	C_{2(ijij)}(\nu) &= \frac{1}{N^2} \int_{\vp,\vp'}  \vp^2 \cdot \vp'^2 \int_{\omega,\omega'}  \, \CG_R(\nu-\omega,-\vp) \CG_R(\omega,\vp) \nonumber \\ 
\fl & \qquad \qquad \times \CG_{\rm eff}(\omega',\omega,\vp',\vp) \CG_R(\nu-\omega',-\vp') \CG_R(\omega',\vp') 
\end{eqnarray}
and 
\begin{eqnarray}
	\label{eq:c1}
\fl	C_{1(ijij)}(\nu) = \frac{1}{N^2} \int_{\vp,\vp'} \vp^2 \left ( \vp \cdot \vp'\right )^2 \vp'^2 \int_{\omega,\omega'} \CG_R(\vp,\omega)\CG_R(-\vp,\nu-\omega) & \nonumber \\
 \times  \CG^W_\lambda(\vp-\vp',\omega-\omega') \CG_R(\vp',\omega')\CG_R(-\vp',\nu-\omega') &
\end{eqnarray}
Or, in full generality:
\begin{eqnarray}
\fl	C_{1(ijij)}(\nu,\vq) &= \frac{1}{N^2} \int_{\vp,\vp'}  (\vq-\vp) \cdot \vp  (\vq-\vp')\cdot \vp' \int_{\omega,\omega'} (\vq-\vp)\cdot(\vq-\vp')  \vp \cdot \vp' \\
\fl	& \times \CG_R(\vp,\omega) \CG_R(-\vp,\nu-\omega) \CG^W_\lambda(\vp-\vp',\omega-\omega') \CG_R(\vp',\omega')\CG_R(-\vp',\nu-\omega') \nonumber \\
\fl	C_{2(ijij)}(\nu) &= \frac{1}{N^2} \int_{\vp,\vp'}  (\vq-\vp)\cdot\vp \,\, (\vq-\vp')\cdot\vp'  \int_{\omega,\omega'}  \, \CG_R(\nu-\omega,-\vp) \CG_R(\omega,\vp) \nonumber \\
\fl & \CG_{\rm eff}(\omega',\omega,\vp',\vp; \vq) \CG_R(\nu-\omega',-\vp') \CG_R(\omega',\vp') 
\end{eqnarray}	
where now
\begin{eqnarray}
\fl	\CG_{\rm eff}(\omega',\omega,\vp',\vp; \nu, \vq) =  \nonumber \\
\fl \quad \int_{\vp'', \omega''} (\vp''-\vp)\cdot \vp \,\, (\vq-\vp)\cdot (\vp''-\vp) \,\, (\vq-\vp')\cdot (\vp'-\vp'') \,\, \vp'  \cdot (\vp'-\vp'') \nonumber \\
\fl	\quad \times  \CG_W(\omega''-\omega,\vp''-\vp) \CG_W(\omega'-\omega'',\vp'-\vp'')   \CG_{R,\lambda}(\nu-\omega'',\vq-\vp'') \CG_{R,\lambda}(\omega'',\vp'')
\end{eqnarray}
{\bf NB.} It really doesn't matter whether $\CG^W_\lambda(\vp'-\vp)$ or $\CG^W_\lambda(\vp-\vp')$ is used. It amounts to a redefinition of momenta that keeps all other terms in $C_1$ the same. Therefore, it must be that $\CG^W_\lambda(\vp'-\vp) = \CG^W_\lambda(\vp-\vp')$

\subsection{Crossed Diagrams}
\label{app:crossed-diagrams}

The first important observation comes from the Fourier space form of a Wightman correlator:
\begin{eqnarray}
	\CG_{W}(\omega,\vq) = \frac{A(\omega,\vq)}{2\sinh \left (\frac{\beta \omega}{2}\right )}
\end{eqnarray}
We see that in real time, $\CG_W(t,\vr)$ has significant support only for $t \sim \beta^{-1}$.
Now consider a ladder diagram of type 1 with two rungs. Its value is, schematically:
\begin{eqnarray}
	\chi_{\rm ladder} &= \int s_1 s_2 s_1' s_2' \CG_R(t-s_1) \CG_R(t-s_1') \CG_R(s_1-s_2) \CG_R(s_1'-s_2') \nonumber \\
	& \times \CG_{W,\lambda}(s_1-s_1') \CG_{W,\lambda}(s_2-s_2') \
	\CG_R(s_2) \CG_R(s_2')
\end{eqnarray}
The retarded Green functions impose, in particular, $\theta(s_1-s_2)\theta(s_1'-s_2')\theta(s_2)\theta(s_2')$, while the Wightman function effectively imposes $s_1 - s_1' \approx 0, s_2 - s_2' \approx 0$. Putting them together the integration phase space is restricted to $\theta(s_1-s_2)\theta(s_2)$.

Now compare this with the case of a crossed diagram
\begin{eqnarray}
	\chi_{\rm crossed} &= \int s_1 s_2 s_1' s_2' \CG_R(t-s_1) \CG_R(t-s_1') \CG_R(s_1-s_2) \CG_R(s_1'-s_2') \nonumber \\
	& \times \CG_{W,\lambda}(s_1-s_2') \CG_{W,\lambda}(s_1'-s_2) \
	\CG_R(s_2) \CG_R(s_2')
\end{eqnarray}
where the constraints are $\theta(s_1-s_2)\theta(s_1'-s_2')\theta(s_2)\theta(s_2')$, coming from the Green functions, and $s_1 - s_2' \approx 0, s_2 - s_1' \approx 0$ from the Wightman function. Together they require $\theta(s_1-s_2)\theta(s_2-s_1)$, to within $\beta^{-1}$. 

That is why we conclude that the crossed diagrams are parametrically suppressed compared to the ladder diagrams, a fact important at late times.

\subsection{Product of retarded Green functions}
\label{app:product-retarded-simplification}

Let us extract the part of Eq.\ref{eq:c1} that contributes to the leading time dependence of $C(t)$. Define $f(\omega) =\CG^W_\lambda(\vp'-\vp,\omega'-\omega)$. We first work with the free expressions for $\CG_R$.

\begin{eqnarray}
	\label{eqn:supplement-simplifying-GR}
	\CG_R(\omega,\vk) &= \frac{1}{(\omega + i0)^2 + \epsilon_\vk^2} = \frac{1}{2\epsilon_\vk} \left [\frac{1}{\omega-\epsilon_\vk + i0} - \frac{1}{\omega+\epsilon_\vk + i0} \right ] \nonumber \\
\fl	\CG_R(\nu-\omega,-\vp)\CG_R(\omega,\vp) &= \frac{1}{4\epsilon_\vp^2} \left [\frac{1}{\nu - \omega -\epsilon_\vp + i0} - \frac{1}{\nu-\omega+\epsilon_\vp + i0} \right ]
		\nonumber \\
		& \qquad \times \left [\frac{1}{\omega-\epsilon_\vp + i0} - \frac{1}{\omega+\epsilon_\vp + i0} \right ] \nonumber \\ 
	C_1(\nu) & \propto I(\omega') =  \int \frac{d\omega}{2\pi}  \CG_R(\vp,\omega)\CG_R(-\vp,\nu-\omega) f(\omega) \nonumber \\
	I(\omega') &= \frac{-i\, f(\omega = \nu - \epsilon_\vp)}{4\epsilon_\vp^2} \left [\frac{1}{\nu-2\epsilon_\vp +i0} - \frac{1}{\nu+i0} \right ] \nonumber \\
	&+ \frac{-i \, f(\omega = \nu + \epsilon_\vp)}{4\epsilon_\vp^2} \left [\frac{1}{\nu+2\epsilon_\vp +i0} - \frac{1}{\nu+i0} \right ] \nonumber
\end{eqnarray}
{\bf NB.} It is important that above $\CG_{W,\lambda}(\omega'-\omega)$ not have any singularities in $\omega$, otherwise we might get extra poles, or worse. 

Now doing the $\omega'$ integral picking up the other poles, at $\omega'=\nu \pm \epsilon_\vp + i0$:
\begin{eqnarray}
\fl	 I_2(\nu,\vp,\vp') &= \int \frac{d\omega'}{2\pi}  \CG_R(\vp',\omega')\CG_R(-\vp',\nu-\omega') I(\omega') \\
	  &= \frac{(-i)^2 \CG_{W,\lambda}(\vp'-\vp, \nu -\epsilon_{\vp'} - \nu + \epsilon_\vp) }{16\epsilon_\vp^2 \epsilon_{\vp'}^2} \nonumber \\
	  & \times \left [\frac{1}{\nu-2\epsilon_\vp +i0} - \frac{1}{\nu+i0} \right ] \left [\frac{1}{\nu-2\epsilon_{\vp'} +i0} - \frac{1}{\nu+i0} \right ] \nonumber\\
	  &+\frac{(-i)^2 \CG_{W,\lambda}(\vp'-\vp, \nu + \epsilon_{\vp'} - \nu - \epsilon_\vp) }{16\epsilon_\vp^2 \epsilon_{\vp'}^2} \nonumber \\
	  & \times \left [\frac{1}{\nu+2\epsilon_\vp +i0} - \frac{1}{\nu+i0} \right ] \left [\frac{1}{\nu+2\epsilon_{\vp'} +i0} - \frac{1}{\nu+i0} \right ]  
\end{eqnarray}

We must now consider the three kinds of poles in $\nu$ that are important for the inverse Laplace transform to $C(t)$: $\nu^{-2}, \nu^{-1}(\nu \pm 2\epsilon_\vp)^{-1}, (\nu \pm 2\epsilon_\vp)^{-1} (\nu \pm 2\epsilon_{\vp'})^{-1}$. 
The $\nu$ integral will be done above all singularities of $I_2(\nu,\vp,\vp')$ in the complex plane
\begin{eqnarray*}
\fl	C(t > 0)_{(1)} & = \frac{-1}{N^2}\int\frac{d\nu}{2\pi} e^{-i\nu t} \frac{1}{(\nu + i0)^2} \int_{\vp,\vp'} \left (\vp \cdot \vp' \right )^4  \\
\fl & \qquad \times \frac{\CG_{W,\lambda}(\vp'-\vp,  \epsilon_{\vp'}- \epsilon_\vp) + \CG_{W,\lambda}(\vp'-\vp,  \epsilon_\vp - \epsilon_{\vp'}) }{16\epsilon_\vp^2 \epsilon_{\vp'}^2} \\
\fl	&= \frac{t}{N^2} \int_{\vp,\vp'} \left (\vp \cdot \vp' \right )^4 \frac{\CG_{W,\lambda}(\vp'-\vp,  \epsilon_{\vp'}- \epsilon_\vp) + \CG_{W,\lambda}(\vp'-\vp,  \epsilon_\vp - \epsilon_{\vp'}) }{16\epsilon_\vp^2 \epsilon_{\vp'}^2} \\
\end{eqnarray*}
and
\begin{eqnarray*}
\fl	C(t > 0)_{(2)} & = \frac{1}{N^2}\int_{\vp,\vp'} \int\frac{d\nu}{2\pi} e^{-i\nu t} \frac{1}{(\nu + i0)(\nu - 2\epsilon_\vp + i0)}  \left (\vp \cdot \vp' \right )^4  \nonumber \\
	& \times \frac{\CG_{W,\lambda}(\vp'-\vp,  \epsilon_\vp - \epsilon_{\vp'}) }{16\epsilon_\vp^2 \epsilon_{\vp'}^2} + \cdots \\
	&= \frac{i}{N^2} \int_{\vp,\vp'} \frac{1+e^{-i 2\epsilon_\vp t}}{2\epsilon_\vp}\left (\vp \cdot \vp' \right )^4 \frac{\CG_{W,\lambda}(\vp'-\vp,  \epsilon_\vp - \epsilon_{\vp'}) }{16\epsilon_\vp^2 \epsilon_{\vp'}^2} + \cdots \nonumber
\end{eqnarray*}
Clearly all poles besides $\nu^{-2}$ give subleading time dependence at late-times, and therefore we will ignore them, and use the simplifying form
\begin{eqnarray}
\fl	\CG_R(\nu-\omega,-\vp)\CG_R(\omega,\vp) &= \frac{\pi i }{2\epsilon_\vp^2}  \left ( \frac{\delta(\omega - \nu + \epsilon_\vp)}{\nu + i0} +  \frac{\delta(\omega - \nu - \epsilon_\vp)}{\nu + i0} \right )
\end{eqnarray}
or because of the two delta functions, might as well
\begin{eqnarray}
	\CG_R(\nu-\omega,-\vp)\CG_R(\omega,\vp) &= \frac{\pi i }{2\epsilon_\vp^2}  \left ( \frac{\delta(\omega + \epsilon_\vp)}{\nu + i0} +  \frac{\delta(\omega - \epsilon_\vp)}{\nu + i0} \right ) \\
	&= \frac{\pi i }{\epsilon_\vp} \frac{\delta(\omega^2 - \epsilon_\vp^2)}{\nu + i0} 
\end{eqnarray}
Finally, anticipating a self-energy induced finite lifetime $\Gamma_\vp = \Im \left [ \Sigma_R(\epsilon_\vp,\vp) \right ]/2\epsilon_\vp$ (see Eqn~\ref{eqn:scattering-rate}):
\begin{eqnarray}
	\CG_R(\nu-\omega,-\vp)\CG_R(\omega,\vp) &= \frac{\pi i }{\epsilon_\vp} \frac{\delta(\omega^2 - \epsilon_\vp^2)}{\nu + i 2 \Gamma_\vp} 
\end{eqnarray}

\subsection{Simplifying $\CG_{\rm eff}$}
To include the contribution of the type 2 rungs we must simplify $\CG_{\rm eff}$ first. In what follows we use the free $\phi$ field expressions for $\CG_W$ based on the observation that the self-energy $\Sigma_R \propto \frac{1}{N}$. In can be verified that:
\begin{eqnarray}
	\CG_{W}(\omega,\vk) &= \CQ(\omega)A(\omega,\vk)\\
	\CQ(\omega) &= \frac{1}{2\sinh\left ( \frac{\beta \omega}{2}\right )} \\
	\CG_W(\omega,\vk) &= Q(\omega) \frac{\pi}{\epsilon_\vk}\left ( \delta(\omega- \epsilon_\vk) - \delta(\omega + \epsilon_\vk) \right )
\end{eqnarray}
Then
\begin{eqnarray*}
\fl	\CG_W^2 & \equiv \CG_W(\omega''-\omega,\vp''-\vp)\CG_W(\omega'-\omega'',\vp'-\vp'') \\
\fl	 &= \frac{\pi^2 \CQ(\omega''-\omega)\CQ(\omega'-\omega'')}{\epsilon_{\vp''-\vp} \epsilon_{\vp'-\vp''}}
	 \left [ \delta(\omega''-\omega-\epsilon_{\vp''-\vp}) - \delta(\omega''-\omega + \epsilon_{\vp''-\vp}) \right ] \\
\fl & \qquad \times	 \left [ \delta(\omega'-\omega''-\epsilon_{\vp'-\vp''}) - \delta(\omega'-\omega'' + \epsilon_{\vp'-\vp''}) \right ] 
\end{eqnarray*}

And after performing the $\omega''$ integral and changing coordinates to $\bar{\vp} = \vp'-\vp, \vP = (\vp+\vp')/2, \bar{\omega} = \omega'-\omega$... we arrive at
\begin{eqnarray}
\fl	\CG_{\rm eff}(\omega',\omega,\bar{\vp},\bar{\vP}) = \frac{1}{2}\int \frac{d^2\vp''}{(2\pi)^2} \left [ (\vP - \frac{\vpb}{2}) \cdot (\vp''+\frac{\vpb}{2}) \,\, (\vP + \frac{\vpb}{2}) \cdot (\vp''-\frac{\vpb}{2})  \right ]^2  \frac{\pi}{\epsilon_+ \epsilon_-}  \\
\fl	 \times  \lbrace Q(\epsilon_+) Q(\bar{\omega}-\epsilon_+) \CG_{R,\lambda}(\nu-\omega-\epsilon_+,-\vp''-\vP) \CG_{R,\lambda}(\omega+\epsilon_+,\vp''+\vP)  \left ( \delta(\bar{\omega}-\epsilon_+ - \epsilon_-) - \delta(\bar{\omega}-\epsilon_+ + \epsilon_-) \right )\nonumber \\
\fl	 - Q(-\epsilon_+) Q(\bar{\omega}+\epsilon_+) \CG_{R,\lambda}(\nu-\omega+\epsilon_+,-\vp''-\vP) \CG_{R,\lambda}(\omega-\epsilon_+,\vp''+\vP)  \left ( \delta(\bar{\omega}-\epsilon_+ - \epsilon_-) - \delta(\bar{\omega}+\epsilon_+ + \epsilon_-) \right )	\nonumber 	
	\rbrace
\end{eqnarray}

Let $\theta$ ($\phi$) be the angle between $\vP$ ($\vp''$) and $\vpb$, then 
\begin{eqnarray}
\fl	\left [ (\vP - \frac{\vpb}{2}) \cdot (\vp''+\frac{\vpb}{2}) \,\, (\vP + \frac{\vpb}{2}) \cdot (\vp''-\frac{\vpb}{2})  \right ]^2  \nonumber \\
\fl \qquad \qquad = \left [ \right (P p'' \cos(\theta-\phi) - \frac{\bar{p}^2}{4}\left )^2  - \left ( \frac{P \bar{p}}{2} \cos\theta - \frac{\bar{p} p''}{2} \cos\phi  \right )^2 \right ]^2
\end{eqnarray}

For $\vq \neq 0$, we can simplify to
\begin{eqnarray}
	\label{eqn:geff-q-simplified}
\fl	\CG_{\rm eff}(\nu,\vq; \omega,\omega',\vp,\vp') = \frac{1}{2}\int \frac{d^2\vp''}{(2\pi)^2} 
	\left[ (\vq -\vP + \frac{\vpb}{2}) \cdot (\vp''+\frac{\vpb}{2}) \,\, (\vq -\vP - \frac{\vpb}{2}) \cdot (\vp''-\frac{\vpb}{2})  \right] \nonumber \\
\fl	 \times \left[ (\vP - \frac{\vpb}{2}) \cdot (\vp''+\frac{\vpb}{2}) \,\, (\vP + \frac{\vpb}{2}) \cdot (\vp''-\frac{\vpb}{2}) \right ] \frac{\pi}{\epsilon_+ \epsilon_-}  \nonumber \\
\fl	\times    \lbrace Q(\epsilon_+) Q(\bar{\omega}-\epsilon_+) 
	\CG_{R,\lambda}(\nu-\omega-\epsilon_+,\vq-\vp''-\vP) \CG_{R,\lambda}(\omega+\epsilon_+,\vp''+\vP)  \left ( \delta(\bar{\omega}-\epsilon_+ - \epsilon_-) - \delta(\bar{\omega}-\epsilon_+ + \epsilon_-) \right )  \nonumber \\	
\fl	- Q(-\epsilon_+) Q(\bar{\omega}+\epsilon_+) \CG_{R,\lambda}(\nu-\omega+\epsilon_+,\vq-\vp''-\vP) 
	\CG_{R,\lambda}(\omega-\epsilon_+,\vp''+\vP)\left ( \delta(\bar{\omega}-\epsilon_+ - \epsilon_-) - \delta(\bar{\omega}+\epsilon_+ + \epsilon_-) \right )		
	\rbrace \nonumber \\
\end{eqnarray}

\section{$\lambda$ self-energy}
\label{app:lambda-self-energy}

\subsection{Zero temperature}
\begin{eqnarray}
\fl	\Pi(i\omega_n,\vq) &= \frac{1}{2}\int \frac{d\nu}{2\pi}\int_\vk^\Lambda \frac{ \left ( \vk \cdot (\vk + \vq) \right )^2  }{\left( \omega_n + \nu \right )^2 + \epsilon^2_{\vk+\vq} } \frac{1}{\nu^2 + \epsilon^2_\vk} \\	
\fl	&= \frac{1}{2}\int_\vk \left ( \vk \cdot (\vk + \vq) \right )^2  \frac{ \epsilon_\vk + \epsilon_{\vk+\vq} }{2 \epsilon_{\vk} \epsilon_{\vk+\vq} \left((\epsilon_\vk + \epsilon_{\vk+\vq})^2+\omega_n^2\right)} \\
\fl	\Pi_R(\omega, \vq) &= \Pi(i\omega = \omega + i0,\vq)\\
\fl	&= \frac{1}{2}\int_\vk \left ( \vk \cdot (\vk + \vq) \right )^2  \frac{1 }{4\epsilon_{\vk} \epsilon_{\vk+\vq} }
	\left [ \frac{1}{( \omega + \epsilon_\vk + \epsilon_{\vq+\vk} + i \epsilon )} -  \frac{1}{(\omega - \epsilon_\vk  - \epsilon_{\vq+\vk} +i \epsilon )}\right ] \\
\fl \mathrm{Im} \, \Pi_R(\omega, \vq) &= \frac{1}{8} \int_\vk  \frac{\left ( \vk \cdot (\vk + \vq) \right )^2  }{\epsilon_{\vk} \epsilon_{\vk+\vq} }
	\left [ -\pi \delta( \omega + \epsilon_\vk + \epsilon_{\vq+\vk} ) + \pi  \delta(\omega - \epsilon_\vk  - \epsilon_{\vq+\vk}) \right ] \\
\fl	&=  \frac{1}{8} \int_\vk  \frac{\left ( (\vk-\vq/2) \cdot (\vk + \vq/2) \right )^2  }{\epsilon_{\vk-\vq/2} \epsilon_{\vk+\vq/2} } \nonumber \\
 &	\times \left [ -\pi \delta( \omega + \epsilon_{\vk-\vq/2} + \epsilon_{\vk+\vq/2} ) + \pi  \delta(\omega - \epsilon_{\vk-\vq/2}  - \epsilon_{\vk+\vq/2}) \right ] 
\end{eqnarray}
where in the last line we shifted $\vk \rightarrow \vk - \vq/2$. Now, some simplifications. We work first with $\epsilon^2_{\vk} = \sqrt{\kappa} \vk^4$.
\begin{eqnarray}
\fl	\epsilon_{\vk+\vq/2}	\epsilon_{\vk-\vq/2} &= \kappa \left (\vk^2 + \frac{\vq^2}{4} + k q \cos\phi \right )\left (\vk^2 + \frac{\vq^2}{4} - k q \cos\phi\right )\\
	&= \kappa  \left [ \left (\vk^2 + \frac{\vq^2}{4} \right )^2 - \vk^2 \vq^2 \cos^2 \phi \right ] \\
\fl	\epsilon_{\vk-\vq/2} + \epsilon_{\vk+\vq/2} &= \sqrt{\kappa} \left ( 2 \vk^2 + \frac{\vq^2}{2}\right )\\
\fl	\left ( (\vk-\vq/2) \cdot (\vk + \vq/2) \right )^2 & = \left ( \vk^2 - \frac{\vq^2}{4}\right )^2
\end{eqnarray}

Since the imaginary part of $\Pi_R$ is easily seen to be odd in $\omega$, we choose to evaluate the $\omega > 0$ part.
\begin{eqnarray}
\fl	\mathrm{Im} \, \Pi_R(\omega > 0, \vq) &= \frac{1}{8} \int_\vk  \frac{\left ( (\vk-\vq/2) \cdot (\vk + \vq/2) \right )^2  }{\epsilon_{\vk-\vq/2} \epsilon_{\vk+\vq/2} }
	 \pi  \delta(\omega - \epsilon_{\vk-\vq/2}  - \epsilon_{\vk+\vq/2}) \\
\fl	 &= \frac{\pi}{8} \int \frac{d\phi k dk}{(2\pi)^2}  \frac{\left ( \vk^2 - \frac{\vq^2}{4}\right )^2  \delta(\omega - \sqrt{\kappa} \left ( 2 \vk^2 + \frac{\vq^2}{2}\right ))  }{ \kappa  \left [ \left (\vk^2 + \frac{\vq^2}{4} \right )^2 - \vk^2 \vq^2 \cos^2 \phi \right ] } 
\end{eqnarray}
For $\omega \geq \sqrt{\kappa} q^2/2$ we have the (physical) root
$$ k_r = +\sqrt{\frac{\omega-\sqrt{\kappa} q^2/2}{2\sqrt{\kappa}}}  = \sqrt{\frac{\omega}{2\sqrt{\kappa}} -\frac{q^2}{4}}$$

Putting it all together, and recalling that $|\vk| \leq \Lambda$:
\begin{eqnarray}
\fl	\mathrm{Im} \, \Pi_R(\omega > 0, \vq) &= \frac{\pi}{8} \int \frac{d\phi }{(2\pi)^2}  \frac{\left ( k_r^2 - \frac{q^2}{4}\right )^2  }{ \kappa  \left [ \left (k_r^2 + \frac{q^2}{4} \right )^2 - k_r^2 q^2 \cos^2 \phi \right ] }
\frac{k_r}{|-4 \sqrt{\kappa} k_r|}  \nonumber \\
 & \times \theta\left ( \Lambda - k_r \right )  \theta \left ( \omega - \frac{\sqrt{\kappa}q^2}{2	}\right ) \\
 &= \frac{\pi}{8} \int \frac{d\phi }{(2\pi)^2}  \frac{\frac{1}{2^2} \left(\frac{\omega }{\sqrt{\kappa }}-q^2\right)^2 }{ \kappa  \left [ \left (k_r^2 + \frac{q^2}{4} \right )^2 - k_r^2 q^2 \cos^2 \phi \right ] }
\frac{1}{4 \sqrt{\kappa}} \\
 &= \frac{\pi}{2^7 \kappa^{3/2}} \left(\frac{\omega }{\sqrt{\kappa }}-q^2\right)^2  \int \frac{d\phi }{(2\pi)^2}  \frac{1}{   \left [ \left (k_r^2 + \frac{q^2}{4} \right )^2 - k_r^2 q^2 \cos^2 \phi \right ] }
\end{eqnarray}
Doing the $\phi$ integral, we have
\begin{eqnarray}
\fl	\mathrm{Im} \, \Pi_R(\omega > 0, \vq) 
 &= \frac{\pi}{32 \kappa^{3/2}} \frac{1}{2^2}   \frac{\left(\frac{\omega }{\sqrt{\kappa }}-q^2\right)^2 }{(2\pi)^2}  \frac{2\pi \theta\left ( \Lambda - k_r \right )  \theta \left ( \omega - \frac{\sqrt{\kappa}q^2}{2	}\right )}{ \sqrt{\left (k_r^2 + \frac{q^2}{4} \right )^2  \left [ \left (k_r^2 + \frac{q^2}{4} \right )^2 -k_r^2 q^2 \right ] }} \\
  &= \frac{\pi}{32 \kappa^{3/2}} \frac{1}{2^2} \left(\frac{\omega }{\sqrt{\kappa }}-q^2\right)^2  \frac{1 }{(2\pi)^2}  \frac{32\pi}{q^4 - 16 k_r^4 }\\
  &= \frac{\pi}{32 \kappa^{3/2}} \frac{1}{2^2} \left(\frac{\omega }{\sqrt{\kappa }}-q^2\right)^2  \frac{1 }{(2\pi)^2}  \frac{8 \pi  \kappa }{\omega  \left(\omega -\sqrt{\kappa } q^2\right)}\\
  &= \frac{\pi}{32 \kappa^{3/2}} \frac{1}{2^2}  \frac{1 }{(2\pi)^2}  \frac{8 \pi    }{\omega } \left(\omega -\sqrt{\kappa } q^2\right)\\
\end{eqnarray}
and finally
\begin{eqnarray}
	\label{eqn:im-pi-zero-temperature}
\fl	 \mathrm{Im} \, \Pi_R(\omega > 0, \vq)  &= \frac{1}{64 \kappa^{3/2}}  \frac{\left(\omega -\sqrt{\kappa } q^2\right) }{\omega } \theta\left ( \omega - \frac{\sqrt{\kappa} q^2}{2}\right ) \theta\left ( 2\sqrt{\kappa}(\Lambda^2 + \frac{q^2}{4}) - \omega \right ) 
\end{eqnarray}

\subsection{Finite temperature}

At $T>0$, we have expressions similar to Appendix B of Chowdhury:
\begin{eqnarray}
\label{eq:impi-finite-temp}
\fl	\Im \left [\Pi_R(\nu + i0 ,q) \right ]_{\nu>0} &= \frac{1}{2} \int_\vk^\Lambda  \frac{\pi \left ( \vk \cdot (\vk + \vq) \right )^2 }{4 \epsilon_\vk \epsilon_{\vk+\vq}} 
	 2 \left [ b(\epsilon_{\vk+\vq}) - b(\epsilon_\vk) \right ]\delta( \nu + \epsilon_{\vk+\vq} - \epsilon_\vk) \nonumber \\
	 + \frac{1}{2} \int_\vk^\Lambda &\frac{\pi \left ( \vk \cdot (\vk + \vq) \right )^2 }{4 \epsilon_\vk \epsilon_{\vk+\vq}} 
	  \left [ b(\epsilon_{\vk+\vq}) - b(-\epsilon_\vk) \right ] \delta(\epsilon_{\vk+\vq} + \epsilon_\vk - \nu )
\end{eqnarray}
So far this expression is general, and $\epsilon_\vk$ could be the one-loop corrected dispersion.
Note that if we set $\vq = 0$ above we obtain:
\begin{eqnarray}
\fl	\Im \left [\Pi_R(\nu + i0 ,q=0) \right ]_{\nu>0} &= \frac{1}{2} \int_\vk \frac{\pi \left ( \vk \right )^4 }{4 \epsilon_\vk^2} 
  \left [ b(\epsilon_{\vk}) - b(-\epsilon_\vk) \right ] \delta(2 \epsilon_{\vk} - \nu )
\end{eqnarray}
This can be finally simplified to:
\begin{eqnarray*}
\fl	\Im \left [\Pi_R(\nu + i0 ,q=0) \right ]_{\nu>0} &= \frac{b(\nu/2)-b(-\nu/2)}{16\nu \sqrt{r^2 + \kappa \nu^2}}\left (\frac{-r + \sqrt{r^2 + \kappa \nu^2}}{2\kappa}\right )^2 \theta \left ( 2\sqrt{\kappa} \Lambda^2 - \nu \right )
\end{eqnarray*}
Now, at $T=0$ we must set $r=0$ and we obtain
\begin{eqnarray}
	\Im \left [\Pi_R(\nu + i0 ,q) \right ]_{\nu>0} &= \frac{1}{64 \kappa^{3/2}} \left ( 2\sqrt{\kappa} \Lambda^2 - \nu \right )
\end{eqnarray}
Sending $\nu \rightarrow 0$ instead we have
\begin{eqnarray}
	\Im  \left[ \Pi_R(\nu,q=0) \right] &= \frac{T}{16\nu^2 r} \frac{\nu^4}{16 r^2} = T\frac{\nu^2}{2^8 r^3}
\end{eqnarray}
which implies that $\CG_{W,\lambda}(\nu,q=0) \rightarrow 0$ as $\nu \rightarrow 0$ because the $1/\sinh(\beta \nu)$ is overcome.

Aftering shifting $\vk \rightarrow \vk - \vq/2$, and $\epsilon_\pm = \epsilon_{\vk + \pm \vq/2}$ in (\ref{eq:impi-finite-temp}):

\begin{eqnarray}
\fl	\Im   \left [\Pi_R(\nu + i0 ,q) \right ]_{\nu>0} &= \frac{1}{2} \int_\vk \frac{\pi \left (  k^2 - q^2/4 \right )^2 }{4 \epsilon_+ \epsilon_-} 
		2 \left [ b(\epsilon_+) - b(\epsilon_-) \right ]\delta( \nu + \epsilon_+ - \epsilon_-) \nonumber \\
		&+ \frac{1}{2} \int_\vk \frac{\pi \left (  k^2 - q^2/4 \right )^2 }{4 \epsilon_+ \epsilon_-}  \left [ b(\epsilon_+)- b(-\epsilon_-) \right ] \delta(\epsilon_+ + \epsilon_- - \nu ) \\
\fl	&= \frac{\pi}{8} \int \frac{d\phi}{(2\pi)^2} \int k \, dk\,  \frac{ \left (  k^2 - q^2/4 \right )^2 }{ \epsilon_+ \epsilon_-} 
	 2 \left [ b(\epsilon_+) - b(\epsilon_-) \right ]\delta( \nu + \epsilon_+ - \epsilon_-) \nonumber \\
	 &+ \frac{\pi}{8} \int \frac{d\phi}{(2\pi)^2} \int k \, dk\,  \frac{ \left (  k^2 - q^2/4 \right )^2 }{ \epsilon_+ \epsilon_-} 
	 \left [ b(\epsilon_+)- b(-\epsilon_-) \right ] \delta(\epsilon_+ + \epsilon_- - \nu ) \nonumber \\
\fl	&= \frac{\pi}{8} \frac{1}{(2\pi)^2} \left ( I_1 + I_2 \right )
\end{eqnarray}

At finite temperature it is imperative that we use the full expression for $\epsilon_\pm^2 = r(T) k_\pm^2 + \kappa k_\pm^4$.
Let us scale somethings out:
\begin{eqnarray}
	A(\nu,q,r,\kappa,T) &= \Im  \Pi_R(\nu,q,r,\kappa,T) \\
			&= B(\frac{\nu}{T}, \frac{q}{\sqrt{T}}, \frac{r}{T},\kappa)
\end{eqnarray}
Since $r(T) = \alpha(T) T$, where $\alpha$ is dimensionless and very weakly dependent on $T$, we will just replace $r/T = \alpha \approx const$, but keeping in mind it is actually a function of temperature.  We can do this once more with dimensionless $\kappa$:
\begin{eqnarray}
	A(\nu,q,r,\kappa,T) &= B(\frac{\nu}{T}, \frac{q}{\sqrt{T}}, \alpha,\kappa) \\
				&= \frac{1}{\kappa^{3/2}}  C(\frac{\nu}{T}, \frac{q \kappa^\frac{1}{4}}{\sqrt{T}}, \frac{\alpha}{\sqrt{\kappa}}) \frac{\pi}{8} \frac{1}{(2\pi)^2}
\end{eqnarray}
where finally
\begin{eqnarray}
	C(\nu,q,\alpha) &= I_1(\nu,q,\alpha) + I_2(\nu,q,\alpha)
\end{eqnarray}
So we need to evaluate two integrals where implicitly we have set $\kappa=1,T=1$ and $\alpha \ll 1$ a constant.
\begin{eqnarray}
\fl	I_1(\nu,q,\alpha) &= 2\int_0^{2\pi} d\phi \int_0^\infty k \, dk\,  \frac{ \left (  k^2 - q^2/4 \right )^2 }{ \epsilon_+ \epsilon_-} 
	\left [ b(\epsilon_+) - b(\epsilon_-) \right ]\delta( \nu + \epsilon_+ - \epsilon_-) \\
\fl	I_2(\nu,q) &= \int d\phi \int k \, dk\,  \frac{ \left (  k^2 - q^2/4 \right )^2 }{ \epsilon_+ \epsilon_-} 
	 \left [ b(\epsilon_+)- b(-\epsilon_-) \right ] \delta(\epsilon_+ + \epsilon_- - \nu )
\end{eqnarray}
Some observations that are useful for testing numerics: $I_2=0$ for $q > q^*$ where $q^*$ is a root of $2\epsilon_\vq = \nu$. 
Recall due the the factor of $\sinh(\beta \omega)$ in $\CG_{W,\lambda}$ we only care about $\omega/T \sim 1$, and certainly not $\omega/T \gg 1$.
We verified numerically that $\Im  \Pi_R(\nu,\vq) \rightarrow \frac{1}{64 \kappa^{3/2}}$ for $\nu \gg T,\kappa \vq^2$, which matches the $T=0,\vq=0$ result.
\subsection{$\Re \Pi_R $}

The KK relation tells us that for a function analytic in the upper half plane (which implies its real-time version vanishes for $t<0$), its real and imaginary parts can be determined from each other.

\begin{eqnarray}
	\Re  \Pi_R(\nu,q) &= \frac{2}{\pi} P \int_0^\infty d\nu \frac{\omega \Im  \Pi_R(\omega,q) - \nu \Im  \Pi_R(\nu,q)}{\omega^2 - \nu^2}
\end{eqnarray}

Armed with $\Re  \Pi_R(\nu,q)$ we have the $O(1)$ dressed $\lambda$ propagator:
\begin{eqnarray}
	\CG_\lambda(i\omega_n,\vq) &= \frac{1}{\frac{-1}{4v}-\Pi(i\omega_n,\vq)} \\
	\CG_{R,\lambda}(\omega+i0,\vq) &= - \CG_\lambda(i\omega_n = \omega + i0,\vq) \\
	 &= \frac{1}{\frac{1}{4v}+ \Pi_R(\omega,\vq)}
\end{eqnarray}

From the $T=0$ expression for $\Im  \Pi_R(\nu,\vq)$ we can see that since $\Im  \Pi_R(\nu,\vq) \rightarrow \frac{1}{64 \kappa^{3/2}}$ at large $\nu$, we expect a logarithmic divergence in the KK transform.

Numerically, it can be handled as follows. Find a number $c$, such that at $\nu = c \nu_*(\vq)$, $\Im  \Pi_R$ is sufficiently close to its asymptotic value. In our case, for $r(T) = 10^{-5} T$, this turns out to be $c \approx 2500$. Split the integral
\begin{eqnarray}
	\Re  \Pi_R(\omega,\vq) &= \frac{2}{\pi} \int_0^{2500 \nu_*(\vq)} d\nu \frac{\omega \Im  \Pi_R(\omega,q) - \nu \Im  \Pi_R(\nu,q)}{\omega^2 - \nu^2} \\
	&+ \frac{2}{\pi} \int_{2500 \nu_*(\vq)}^E d\nu \frac{\omega \Im  \Pi_R(\omega,q) - \nu \Im  \Pi_R(\nu,q)}{\omega^2 - \nu^2}
\end{eqnarray}
The first line can be done numerically, while the second can be done analytically with the numerics-backed observation that the imaginary part of $\Pi_R$ doesn't change much.
\begin{eqnarray}
\Re  \Pi_R(\omega,\vq) &= \frac{2}{\pi} \int_0^{2500 \nu_*(\vq)} d\nu \frac{\omega \Im  \Pi_R(\omega,q) - \nu \Im  \Pi_R(\nu,q)}{\omega^2 - \nu^2} \\
-& \frac{2}{\pi} \Im  \Pi_R(\omega,\vq) {\rm Arctanh}\left (\frac{\omega}{2500 \nu_*(\vq)}\right ) \\
&+ \frac{1}{64\pi \kappa^{3/2}} \log \left | \frac{E^2 - \omega^2}{(2500 \nu_*(\vq))^2 - \omega^2}\right |
\end{eqnarray}

In subsequent computations we drop the term in the last line above.

\section{Scattering rate of $\phi$}
\label{app:phi-scattering-rate}

We need to evaluate $G_R(\omega,\vq)$, and more specifically
\begin{eqnarray}
	\Sigma\left (i\omega_n,\vq \right ) &= \frac{T}{N}\sum_{ i \nu_n} \int_\vk^\Lambda \left ( \vq \cdot (\vq+\vk) \right )^2 \CG(i\omega_n + i\nu_n, \vq+\vk) \CG_\lambda(i\nu_n,\vk)
\end{eqnarray}

Following the Chowdhury\cite{ChowdhurySwingle2017} we derive
\begin{eqnarray}
\fl	\Im  \Sigma_R(\omega + i0,\vq) &= \frac{1}{N} \int_\vk^\Lambda \left ( \vq \cdot \vk \right )^2  \frac{\sinh(\beta \omega/2)}{4\epsilon_\vk \sinh(\beta \epsilon_\vk)/2} \nonumber \\
 & \qquad \times \left [ \CG_{W,\lambda}(\epsilon_\vk-\omega,\vk-\vq) + \CG_{W,\lambda}(-\epsilon_\vk-\omega,\vk-\vq) \right ]
\end{eqnarray}

Written in rescaled variables
\begin{eqnarray}
	\label{eqn:im-phi-rescaled}
\fl	\Im  \Sigma_R(\omega T,\vq \sqrt{T}) &= \frac{T^2}{N} \int_\vk^{\Lambda/\sqrt{T}}\left ( \vq \cdot \vk \right )^2  \frac{\sinh( \omega/2)}{4\epsilon_\vk \sinh( \epsilon_\vk)/2} \nonumber \\
 & \qquad \times \left [ \CG_{W,\lambda}(\epsilon_\vk-\omega,\vk-\vq) + \CG_{W,\lambda}(-\epsilon_\vk-\omega,\vk-\vq) \right ]
\end{eqnarray}

Some observations: At $T=0$, $\Im \Pi_R$ is cutoff independent (see Eq.\ref{eqn:im-pi-zero-temperature}), but $\Re \Pi_R$ is clearly cutoff dependent and goes like $\log \Lambda$ because $\Im \Pi_R(\nu,\vq) \rightarrow \frac{1}{64 \kappa^{3/2}}$ as $\nu \rightarrow \infty$. Since the location of the pole is a physical quantity, we imagine adding corresponding counterterms that keep the $\nu = \sqrt{r(T) \vk^2 + \vk^4}$ dispersion for $\phi$ once $\Pi$ is used to compute corrections to $\CG$. Therefore, the theory can be fixed such that $\CG_{W,\lambda}$ is cutoff independent.

Proceeding further, we note that in Eq.\ref{eqn:im-phi-rescaled}, the cutoff $\Lambda$ can be safely sent to infinity.

Finally, assuming $\Sigma_R(\epsilon_\vq,\vq)$ is smaller than $\epsilon_\vq$ (always true for sufficiently large $N$) we define the inverse lifetime
\begin{eqnarray}
\fl	\Gamma_\vq &= \frac{\Im  \Sigma_R(\epsilon_\vq,\vq)}{2\epsilon_\vq}\\
	\label{eqn:scattering-rate}
\fl	\Gamma(\vq \sqrt{T} ) &= \frac{T}{2N} \int_\vk^{\Lambda/\sqrt{T}}\left ( \vq \cdot \vk \right )^2  \frac{\sinh( \omega/2)}{4 \epsilon_\vq \epsilon_\vk \sinh( \epsilon_\vk)/2} \nonumber \\ 
 & \qquad \times \left [ \CG_{W,\lambda}(\epsilon_\vk-\omega,\vk-\vq) + \CG_{W,\lambda}(-\epsilon_\vk-\omega,\vk-\vq) \right ]
\end{eqnarray}

\bibliographystyle{unsrt}
\bibliography{LifshitzChaos}

\end{document}